\nonstopmode
\documentstyle[12pt,aasms]{article}
\def\gapprox{\mathrel{\vcenter{\offinterlineskip \hbox{$>$}
    \kern 0.3ex \hbox{$\sim$}}}}
\def\lapprox{\mathrel{\vcenter{\offinterlineskip \hbox{$<$}
    \kern 0.3ex \hbox{$\sim$}}}}
\def\refindent{\par\penalty-100\noindent\parskip=4pt plus1pt
               \hangindent=3pc\hangafter=1\null}
\def\half{{1\over 2}}
\begin{document}
\setlength{\baselineskip}{12pt}

\title{Nonlinear Evolution of the Magnetorotational Instability in
Ion-Neutral Disks}
\author{John F. Hawley}
\affil{Department of Astronomy, University of Virginia, \\
Charlottesville, VA 22903; jh8h@virginia.edu}
\author{James M. Stone}
\affil{Department of Astronomy, University of Maryland, \\
College Park, MD 20742; jstone@astro.umd.edu}

\begin{abstract}
We carry out three-dimensional magnetohydrodynamical simulations of the
magnetorotational (Balbus-Hawley) instability in weakly-ionized
plasmas.  We adopt a formulation in which the ions and neutrals each
are treated as separate fluids coupled only through a collisional drag
term.  Ionization and recombination processes are not considered.  The
linear stability of the ion-neutral system has been previously
considered by Blaes \& Balbus (1994).  Here we extend their results
into the nonlinear regime by computing the evolution of Keplerian
angular momentum distribution in the local shearing box approximation.
We find significant turbulence and angular momentum transport
when the collisional frequency is on order 100 times the orbital 
frequency $\Omega$.  At higher collision rates, the two-fluid
system studied here behaves much like the fully ionized systems studied
previously.  At lower collision rates the evolution of the instability
is determined primarily by the properties of the ions, with the
neutrals acting as a sink for the turbulence.  Since in this regime
saturation occurs when the magnetic field is superthermal with respect
to the ion pressure, we find the amplitude of the magnetic energy and
the corresponding angular momentum transport rate is proportional to
the ion density.  Our calculations show
the ions and neutrals are essentially decoupled when the collision
frequency is less than $0.01\Omega$; in this case the ion fluid behaves
as in the single fluid simulations and the neutrals remain
quiescent.  We find that purely toroidal initial magnetic field
configurations are unstable to the magnetorotational instability across
the range of coupling frequencies.

\end{abstract}

\section{Introduction}

The key to understanding accretion disk dynamics lies with the angular
momentum transport mechanism.  Since the molecular viscosity of disks
is very low, some form of ``anomalous viscosity'' must be present.
Although the precise nature of this anomalous viscosity has long been
elusive, the discovery that differentially rotating systems are
magnetohydrodynamically (MHD) unstable (Balbus \& Hawley 1991) has led
to the conclusion that fully ionized disks must be MHD turbulent.
Because this magnetorotational instability is caused by angular
momentum transport, the resulting turbulence has precisely the right
character to transport angular momentum outwards (Balbus, Hawley \&
Stone 1996, hereafter BHS96), as required for disks to accrete.  Since
the MHD instability plays such a fundamental role in disks, the
question naturally arises, what is its behavior when the plasma is not
fully ionized?  Protostellar and protoplanetary disks as well as other
molecular disks are the venues where this question is particularly
significant.

Although it is tempting to assume that in the absence of MHD
turbulence, purely hydrodynamical turbulence will rise to the task of
transporting angular momentum, this now seems highly unlikely.  BHS96
demonstrated, through a combination of analysis and simulation, that
differentially rotating systems are both linearly and nonlinearly
locally stable to hydrodynamic perturbations so long as the standard
Rayleigh criterion is satisfied.  Even if initiated ``by hand,''
hydrodynamic turbulence is not self-sustaining.  Outward  transport of
angular momentum through a net Reynolds stress requires a specific
average correlation between the radial velocity and the angular
momentum fluctuations in the turbulent flow.  Of fundamental importance
is the interaction of these velocity fluctuations with the background
mean flow.  In differentially rotating systems the source of free
energy is the angular velocity gradient.  Angular momentum
fluctuations, on the other hand, act to reduce the background angular
{\it momentum} gradient, which has the opposite sign from the angular
velocity gradient.  A positive value of the Reynolds stress, required
to tap into the free energy of the system, acts as a sink term for the
evolution of the angular velocity fluctuations that make up the
Reynolds stress itself.  Thus the turbulence is not self-sustaining.

The results of BHS96 further imply that enhanced angular momentum
transport is not the necessary outcome of turbulence.  Because
transport requires a high degree of correlation between the radial and
azimuthal velocity fluctuations, turbulence that does not have its
origin in the mean differentially rotating flow is unlikely to be an
efficient source of angular momentum transport.  This point is
emphasized by Stone \& Balbus (1996), whose numerical simulation showed
that while vertical convection can generate turbulence, the resulting
net radial angular momentum transport was {\it inward} at a low rate
(so long as the equator was kept hot by the boundary conditions of the
simulation).  Similar results were found by Cabot (1996).  Simulations
by Ryu \& Goodman (1994) of the action of a parametric tidal
instability on a disk show the generation of turbulence, but without
internal transport.  These studies provide compelling evidence that the
idea of purely hydrodynamic turbulence as an {\it angular momentum
transport mechanism} in protostellar disks should be discarded.  If
turbulence is to be driven by the differential rotation, it must be MHD
turbulence.  If hydrodynamic turbulence is present, it must be driven
by some source other than the background shear and generally will not
transport angular momentum.  Such considerations are particularly
important if turbulence is important in protoplanetary disks for
chondrule and planetesimal formation (e.g., Cuzzi, Dobrovolskis \&
Hogan 1996).

It now appears that the number of different mechanisms for transporting
angular momentum in disks is very limited.  Magnetic fields and spiral
wave disturbances (nonaxisymmetric distortions) are the only
viable general mechanisms.  In the latter category, and within the
context of protostellar disks, nonaxisymmetric gravitational
instabilities may be important.  Global wave mechanisms, however,
are not normally associated with the generation of local turbulence.
The dynamics of a disk controlled by such global waves will be
quite different from a model based upon the usual local $\alpha$
prescription for viscosity, which is itself based upon the ansatz of
transport by local turbulent stresses.

For protostellar disks we are forced to
reckon with questions of the effectiveness of magnetic field coupling.
There is, of course, nothing about this issue that is unique to the
magnetorotational instability.  If the magnetic field is not well
coupled to the fluid, then all potential MHD processes in the disk will
be affected.  It is difficult to imagine, for example, a scenario in
which a weak field in a disk is stablized by low ion-neutral coupling,
yet remains involved in, say, a kinematic dynamo.  If a protostellar
disk or rotating molecular cloud is sufficiently well-coupled to a weak
magnetic field for the field to be important in any way, then the
magnetorotational instability is an important dynamical
factor.

Under what circumstances, then, will the instability operate and
produce MHD turbulence in protoplanetary disks where the ionization
fraction is quite low?  The first step towards answering this question
was taken by Blaes \& Balbus (1994, hereafter BB), who performed an
axisymmetric linear stability analysis for a vertical (and toroidal)
field in a weakly ionized plasma for a number of limits.  Their major
finding is that the linear magnetorotational instability is present so
long as the ion-neutral collision frequency exceeds the local epicyclic
frequency.  This condition will be satisfied even for very small
ionization fractions.  They also found that azimuthal fields can reduce
the observed growth rates, although such fields do not eliminate the
instability.

While a linear analysis can indicate the presence or absence of the
instability, its effectiveness as a transport mechanism must be
determined by its nonlinear evolution.  The first numerical study of
the two-fluid magnetorotational instability was carried out by Mac Low
et al. (1995).   They assume ionization-recombination equilibrium and
consider the low-ionization limit, neglecting the ion pressure and
inertia.  This reduces the problem to a single-fluid (neutral) plus a
diffusion term in the induction equation (ambipolar diffusion limit).
Using the ZEUS code they carried out a series of two-dimensional
simulations of the vertical flux tube problem (as in Hawley \& Balbus
1991) for various ion-neutral coupling strengths.  Although these
simulations did not follow the evolution much beyond the linear stage,
the results were in agreement with the stability analysis of BB in the
appropriate limit.  Essentially, so long as there are unstable
wavelengths available, and the coupling between ions and neutrals is
sufficiently strong, the instability behaves much like the single fluid
case.  The instability ceases to operate when the ambipolar diffusion
rate becomes comparable to the growth rate of the instability, i.e.,
the field diffusion time is $< \Omega^{-1}$, where $\Omega$ is the
disk orbital frequency.

In another study,
Brandenburg et al. (1995) investigated the ambipolar diffusion limit in 
three-dimensional simulations of a local, vertically
stratified disk.  They considered a case where the ambipolar diffusion
time was long compared to the orbital time.  They
found that in this limit (i.e., ambipolar diffusion sufficiently small)
the instability remains effective, and continues to generate
self-sustained turbulence that transports angular momentum outward,
albeit at a rate slightly reduced from the fully-coupled case.  In
another simulation, the diffusion time was set comparable to
$\Omega^{-1}$ and the turbulence decayed.

These first results, while important, are only a beginning.  To
date, all the numerical studies have considered the behavior of the
partially ionized system in the limit where the inertia of the ions can
be completely neglected, and where the ion density is everywhere a
fixed power-law function of the neutral density.  In this paper we
will approach the problem using a genuine two-fluid, ion-neutral
evolution to examine effects where the ions are free to move relative
to the neutrals.  We will investigate the transition from the
well-coupled regime, through critical coupling where the collision
frequency is comparable to the epicyclic frequency, and down to the
fully uncoupled limit.  This parameter study should more clearly define
the physical conditions for which full MHD turbulence and accompanying
angular momentum transport can be expected, and those for which the
bulk of the system dynamics are essentially hydrodynamic.   The full
range of conditions is likely to be of importance somewhere within
protostellar or protoplanetary disks.  In some regions of these disk
systems, the ionization fractions may be quite small, but in other
regions, such as near the forming protostar, the temperatures will be
high, and the gas will be nearly fully ionized.  In between, there will
be a transition region.   Determining the size of this transition
region, and delineating its properties, will depend on obtaining a
better understanding of the nonlinear behavior of the ion-neutral
system in various regimes.

The plan of the paper is as follows.  In \S2 we will consider the
equations and the numerical techniques used to solve them.   Although
the collision term is handled semi-implicitly, the use of explicit
finite-differencing for the remainder of the system makes the code
Courant-limited.  As a test problem, we compare numerical growth rates
with analytic values from BB.  In \S3 we present the results of an
extensive ensemble of simulations,  covering a range of ionization
fractions and coupling frequencies.  Because of the Courant limit, this
study is limited to relatively large ionization fractions $f \equiv
\rho_i/(\rho_i + \rho_n)$.  In the linear limit, however, the growth
rates for small ionization fractions are relatively unaffected by
decreasing $f$, if the ratio of the coupling frequency to orbital
frequency is held constant.  Thus, we expect that physical insights
gained by this study should extend even into the small $f$ regime.  The
implications of the simulations will be summarized and discussed in \S4.

\section{Two-Fluid Algorithm}

\subsection{Equations and Numerics}

In these simulations we will be studying the simple two-fluid system as
described in \S 3 of BB,  consisting of separate compressible ion and
neutral fluids, coupled only by collisions.  The ion fluid is assumed
to be perfectly conducting.  We include no ionization or recombination;
ions and neutrals are separately conserved.  We have examined both
isothermal systems, and ones where the two fluids can have different
temperatures and no thermal coupling is included.

As in previous work (Hawley, Gammie \& Balbus 1995, hereafter HGB95) we
restrict our attention to the local Hill (1878) system representation of a
disk.  This model incorporates Coriolis and tidal forces, and is
constructed by expanding the equations of motion in cylindrical
coordinates $(R,\phi, z)$ to first order around a fiducial
radius $R_{\circ}$ using a set of local Cartesian coordinates ${\bf x}
= (x,y,z) = (R-R_{\circ}, R_{\circ}[\phi-\Omega t], z)$.  The rotation
law is assumed to have the form $\Omega \sim R^{-q}$; for a Keplerian
disk $q=3/2$.  The resulting equations for the ions, neutrals, and
magnetic field are
\begin{equation}
{\partial\rho_i\over{\partial t}} + \nabla \cdot (\rho_i {\bf v}_i) = 0,
\end{equation}
\begin{equation}
{\partial\rho_n\over{\partial t}} + \nabla \cdot (\rho_n {\bf v}_n) = 0,
\end{equation}
\begin{equation}
{\partial {\bf v}_i\over{\partial t}} + {\bf v}_i\cdot \nabla {\bf v}_i
= - {1\over\rho_i}\nabla \left( P_i + {B^2\over{8 \pi}}\right)
        + {({\bf B}\cdot\nabla){\bf B} \over{4 \pi\rho_i}}
        - 2 {\bf\Omega} \times {\bf v}_i
        + 2 q \Omega^2 x {\hat{\bf x}}
	+ \gamma\rho_n ({\bf v}_n - {\bf v}_i),
\end{equation}
\begin{equation}
{\partial {\bf v}_n\over{\partial t}} + {\bf v}_n\cdot \nabla {\bf v}_n
= - {1\over\rho_n}\nabla  P_n 
        - 2 {\bf\Omega} \times {\bf v}_n
        + 2 q \Omega^2 x {\hat{\bf x}}
	- \gamma\rho_i ({\bf v}_n - {\bf v}_i),
\end{equation}
\begin{equation}
{\partial {\bf B}\over{\partial t}} = \nabla \times 
({\bf v}_i \times {\bf B} ),
\end{equation}
where the terms have their usual meaning, the subscripts $n$ and $i$
refer to the neutrals and ions respectively, and $\gamma$ is the drag
coefficient due to collisions between the two fluids.  In addition,
there is an equation of state (we use either an adiabatic or an
isothermal equation of state) for the neutrals and for the ions.  In
these equations we neglect vertical gravity and assume that the ion
component is a perfectly conducting fluid.  While there are physically
important regimes where resistivity will be important (see, e.g.,
the discussion in BB), we make this assumption
to simplify the parameters of the present study.
The results of this paper will correspond to the most favorable conditions
for the MHD instability since finite resistivity would be
expected to reduce the levels of the resulting turbulence.

Equations (1)--(5) amount to separate hydrodynamical (neutrals) and MHD
(ions) systems, coupled by the collisional drag term.  We use the
hydrodynamical algorithms described by Stone \& Norman (1992a) to
evolve the equations for the neutrals (eqs.  [1--2] and
[4]), and the MOCCT algorithm (Stone \& Norman 1992b; Hawley \& Stone
1995) to evolve the ions (eqs. [3] and [5]).  The collisional drag
term $\gamma$ in both equations (3) and (4) is operator-split
and updated using fully implicit
backward Euler differencing (a complete description of the resulting
difference equations is given in Stone 1997).  This avoids the
Courant stability criterion associated with diffusion of the magnetic
field that restricts explicit differencing formalisms of the drag
term.  The resulting hybrid hydrodynamical-MHD code has been tested
with steady C-type shock solutions and comparison to the analytic
growth rate of the Wardle instability (Stone 1997).  We describe a
comparison of the numerical growth rate of the 
magnetorotational instability in a
partially ionized disk with the analytic value from BB in \S2.2  below.

The computations are done in the periodic three-dimensional (3D)
shearing box (HGB95).  In this system  the computational domain is a
rectangular box with sides $L_x, L_y,$ and $L_z$, and it is assumed to
be surrounded by identical boxes that are strictly periodic at $t =
0$.  A large-scale continuous linear shear flow is present across all
the boxes.  At later times the computational box remains periodic in
$y$ and $z$, while the radial ($x$) boundary condition is determined by
the location of the neighboring boxes as they move relative to one
another due to the background shear.  We use box dimensions of
$L_x=1$, $L_y = 2\pi$, and $L_z = 1$, as in previous single-fluid
simulations (HGB95).  

The initial equilibrium system has a Keplerian
rotation law ($q=3/2$); the angular speed at the center of the shearing
box is $\Omega$.  The initial state has constant densities and
pressures (ion and neutral).  For most experiments we use an isothermal
equation of state; in a few simulations we use of an adiabatic equation
of state with $P\propto \rho^{5/3}$.  The mean molecular weight of the
ions and the neutrals can be specified separately; for most simulations
we adopt the weights used in BB, $m_n = 2.33m_H$ and $m_i = 30 m_H$.
These numbers come from assuming that the neutrals are hydrogen and
helium molecules while the ions correspond to trace alkali species
(Draine, Roberge, \& Dalgarno 1983).  Assuming thermal equilibrium, the
ratio of the initial sound speeds in the neutrals and ions is equal to
the square root of the ratio of the mean molecular weights.

One of the difficulties of this study is the wide range of possible
physical parameters that could be investigated.  Among these are the
equation of state, neutral to ion mass ratio, initial field strength
and orientation, the initial ionization fraction $f \equiv
\rho_i/(\rho_i + \rho_n)$, and the collisional frequency,
$\rho_i\gamma/\Omega$.  Our primary focus will be on the role of the
collisional drag term $\gamma$.  In the limit of $\gamma \rightarrow 0$,
the ion and neutral fluids decouple; the ions evolve as in the
magnetized single-fluid case, while the neutrals evolve
hydrodynamically.  In the other extreme, $\gamma \rightarrow \infty$,
the system reduces to a single well-coupled fluid.  In this study, we
will examine a range of collisional frequencies on either side of the
critical frequency, $\gamma\rho_{i} \sim \Omega$, by treating
$\gamma$ as a free parameter and varying it for
given values of ionization fraction and $\Omega$.  The
Courant condition from the ion Alfv\'en speed limits the simulations to
moderate values of the ionization fraction $f \ge 0.001$. As BB have
shown, it is the ratio of the collisional frequency to the orbital
frequency that determines the linear behavior of the instability.
Assuming this holds even into the nonlinear regime, our results should
be able to set important limits even for significantly smaller, and
hence more realistic, ionization fractions.

\subsection{Two-Dimensional Test Simulation } 

As a simple verification test of the two-fluid code, we compare
observed growth rates in the linear regime of the instability with
analytic values from BB.  (Similar tests were performed for the
fully-coupled, single-fluid system by Hawley and Balbus 1992.) We
choose an initial magnetic field strength corresponding to $\beta_{i} =
P_{ion}/P_{mag} = 2$, run several models with  $f=0.1$ and different
values $\gamma$, and compare the observed growth rate in the perturbed
magnetic radial field $B_x$ with the value given by the dispersion
relation (BB eq. [25]) for the largest vertical wavelength permitted
by the computational domain.  The particular initial magnetic field
strength used makes the (analytically determined) fastest growing
wavelength close to the domain size $L_z=1$.  We use two resolutions,
$63\times 63$ and $31\times 31$ grid zones; these are the $(x,z)$
resolutions that will be used in the 3D simulations to follow.  The
numerical growth rates, in units of $\Omega$, are plotted on top of an
analytic growth rate curve obtained from equation (24) of BB (Fig. 1).
As can be seen from the figure, the two-fluid code reproduces the
analytic linear growth rates for the range of collision frequencies
considered.  There is not much difference between the high and low
resolution simulations; the long wavelength modes considered here are
adequately resolved by either grid.

\section{Results}

\subsection{A Review of Single Fluid Results} 

In this paper our primary focus will be on
an initially uniform vertical magnetic field, $B_z$.  The
single-fluid, fully-ionized version of this problem was investigated by
HGB95, and we begin with a brief summary of their results.  

The maximum growth rate of the single-fluid system is $0.75 \Omega$.
HGB95 found this growth rate for wavenumbers $kv_{Az}/\Omega \simeq 1$ in
all simulations where the fastest growing wavelength is adequately
resolved on the grid.  The fastest growing mode is axisymmetric and
leads to flow along radial ``channels.''  Outward flowing channels
contain excess angular momentum, while inward flowing channels have
less than the Keplerian value.  When the fastest growing mode has a
vertical wavelength greater than or equal to the radial dimension of
the box, this axisymmetric mode continues to grow exponentially well
into the strong field regime.  However, for smaller vertical
wavelengths (larger vertical wavenumbers), the channels break down into
turbulence.  These numerical results can be understood through the
analysis of Goodman \& Xu (1994) who found that, in the local limit and
for a uniform vertical field, the exponentially-growing linear solution
is also an exact {\it nonlinear} solution.  Goodman \& Xu further carried out
a stability analysis of the streaming channel solution and found that
it was subject to a number of instabilities.  The most important of
these parasitic instabilities is a magnetized Kelvin-Helmholtz mode
that is present for radial wavelengths larger than the channel
solution's vertical wavelength.  Thus, under most circumstances, the
channel solution is unstable, and the general outcome of the
instability is turbulence.

The resulting turbulence is nonisotropic; correlated fluctuations in
the magnetic and velocity fields transfer angular momentum outward
through the action of the Maxwell (magnetic) and the Reynolds
stresses.  Accretion disk models often make use of the Shakura \& Sunyaev
(1973) parameterization for the stress $W_{r\phi}$, scaling it with the
total pressure, $W_{r\phi} = \alpha P$.  In the MHD turbulence,
however, the level of net transport depends primarily upon
the magnetic field levels in the saturated turbulent state.
The stress is proportional to the {\it magnetic} pressure with
a ratio of the stress to $P_{\rm mag}$ of about 0.6.  We refer to this
ratio as $\alpha_{\rm mag}$.  

The complete ensemble of HGB95 vertical field simulations provides an
empirical relation for the average magnetic field energy in the
turbulence:  it is proportional to the product of the fastest growing
initial wavelength and the maximum possible unstable wavelength
permitted in the computational domain.  From HGB95, we have
\begin{equation}
\langle {B^2\over 8\pi} \rangle \propto \rho (L_z\Omega)(\lambda_c\Omega),
\end{equation}
where $\lambda_c$ is the fastest growing wavelength for the initial
magnetic field ($\lambda_c = 2\pi [16/15]^{1/2} v_A / \Omega$).  A best
fit to the HGB95 simulations gives a constant of proportionality of
1.2.  In essence, the final energy is the geometric average of the
energy of the {\it net} vertical field piercing the box (which, because
of the periodic boundary conditions does not change as a result of the
evolution) and the energy of the largest magnetic field that would be
unstable given the dimension $L_z$.  The box dimension $L_z$
was set equal to $c_s/\Omega$ to represent a disk scale height,  
although vertical gravity was not included in these simulations. 
The growth of the instability and its
saturation level are unaffected by the hydrodynamic pressure in the
box;  the magnetic field energy remains subthermal (since the 
dimensions of
the computational domain were chosen so that the sound speed was equal
to $L_z\Omega$.)   The saturation amplitude is also unaffected by the
presence of a subthermal toroidal field except to the extent that such
a toroidal field and its instabilities add to the total magnetic
energy.

\subsection{Two-fluid system:  uniform $B_z$ fields} 

We now turn to the two-fluid system and begin with a simple initial
field geometry, a uniform $B_z$ field; in subsequent sections we
consider the evolution of $B_z$ fields with zero net flux, uniform
$B_y$ fields, and mixtures of these cases.

In the two-fluid systems there are two identifiable limits in which the
single-fluid results should apply:  $\gamma = 0$ and $\gamma
\rightarrow \infty$.  When $\gamma = 0$ the ions are completely
decoupled from the neutrals and their evolution will be as a single
fluid with the ion density and ion Alfv\'en speed determining the final
magnetic energy levels.  In the other limit, the system will again
behave as a single fluid, but with the {\it total} density determining
the Alfv\'en speed and saturation energies.  BB point out that the
linear properties of the two-fluid system can be understood in terms of
the ``effective'' density of the system, which varies between $\rho_i$
and $\rho_i + \rho_n$ depending on the strength of the coupling.  Since
$\lambda_c \propto v_A = B/4\pi\rho^{1/2}$, the scaling relation (6)
implies that the ratio of the magnetic energy in the decoupled
($\gamma=0$) limit to the magnetic energy in the fully coupled
($\gamma=\infty$) limit will be $f^{1/2}$, for fixed initial magnetic
field strength.  We will explore the properties of the transition
region between these two limits by varying the drag coefficient
$\gamma$, the ionization fraction $f$, and other variables such as the
equation of state and initial field strength in a number of different
models.

\subsubsection{A fiducial run} 

We begin our discussion with a baseline simulation against which to
compare other models.  This fiducial run (Z17) is a two-fluid shearing
box with $31\times 63\times 31$ resolution, the standard survey
resolution.  (The complete list of vertical field simulations is
provided in Table 1.) We set the neutral density
$\rho_n =1$, and the neutral sound speed $c_s = L_z\Omega$, with $L_z
= 1$ and $\Omega = 10^{-3}$.  The ion
fraction is set to $f=0.1$.  This, together with a the drag coefficient
$\gamma = 0.01$, yields a coupling frequency of $\gamma \rho_i/\Omega =
1.11$, just above the critical value.  We use an isothermal equation of
state, and assume a ratio of ion to neutral mass $m_{i}/m_{n}$ of
$30/2.33 =12.9$.  The ion sound speed is equal to $c_{si} =
(m_{i}/m_{n})^{1/2}$.  The assumption that the ions and neutrals have
the same temperature establishes the relative pressures.  The initial
field is a uniform $B_z$ field with strength $\beta_{i} = 80$, which
corresponds to a value of $kv_{A{i}}/\Omega = 0.28 L_z/\lambda$.  From
the linear dispersion relation [eq. (24) of
BB] we find that the unstable growth rates increase with increasing
wavenumber, from a rate of $0.15\Omega$ for $\lambda=L_z$ to
$0.43\Omega$ for $\lambda = L_z/4$.  Smaller wavelengths are also
unstable with comparable growth rates, up to the limit of $\lambda =
0.068 L_z$, but such wavelengths are not well resolved.

The uniform initial conditions are perturbed with small, random
fluctuations in the ion and neutral pressures.  The magnetic field
energy grows exponentially for the first three orbits before leveling
out after orbit 4 (Fig.  2).  The numerical growth rate of the
instability is computed from the slope of the radial field energy
density curve.  The radial field perturbation grows at a rate of
$0.27\Omega$, equal to the analytic value for $\lambda=L_z/2$.  At 4.6
orbits plots of variables such as the $B_x$ magnetic field show
significant $\lambda = L_z/4$ structure.  At this point, the system is
largely axisymmetric, having the appearance of the channel solutions
previously seen in pure $B_z$ single fluid simulations (Hawley \&
Balbus 1992; HGB95).

At orbit 15 the toroidal field is dominated by a nearly axisymmetric
$\lambda =L_z$ structure.  The ions are greatly affected by the field
since the toroidal field is now superthermal with $\beta_{i} = 0.6$.
Filaments of high density ions  are sandwiched between these regions of
strong toroidal field (Fig. 3).  The maximum ion density in the
filaments is 2.5 times the average value.  
Since the neutral density has a much smaller
relative variation, the ionization fraction in the filaments has gone
up by a comparable amount.  The neutral density plot shows much less
small scale structure; the neutrals are dominated by trailing $m=1$
pressure waves that are nearly independent of $z$.  These have a
maximum $\delta P_{n}/P_{n}$ of about 10\%.  Thus considerable
separation of the ions from the neutrals occurs when the coupling
frequency is comparable to the orbital frequency.

Figure 2 shows significant magnetic energy fluctuations  
after saturation.  Representative values are
obtained by time and space averages; some of these values are listed in
Table 2.  The time-averaged magnetic field energy after saturation is
$B_y^2/8\pi = 0.011 (L_z\Omega)^2$; 
the majority of the energy is in the toroidal field
with $B_x^2/B_y^2 = 0.048$, and $B_z^2/B_y^2 = 0.035$.  Since the toroidal
field grows by both the action of the instability and the
background shear flow, the amplitude of the poloidal fields provides
the best measure of the turbulence.  In this simulation the poloidal fields
are considerably smaller than in single fluid runs.  For example, in
the fiducial run of HGB95 $B_x^2/B_y^2 = 0.32$, and $B_z^2/B_y^2 =
0.10$.  Interestingly, the average magnetic field energy in the
two-fluid system lies below both of the two single-fluid limits given
by equation (6), either the one appropriate for the ions alone, or
the one appropriate to the total density of ions plus neutrals.  This
implies that near the critical coupling frequency  the neutrals
primarily act as a drag to reduce the vigor of the field amplification,
the  MHD turbulence and the resulting transport.

The time-averaged perturbed kinetic energies in the ions are $\half
\langle \rho v_x^2\rangle_i = 6.8\times 10^{-4}(L_z\Omega)^2$, $\half
\langle \rho \delta v_y^2\rangle_i = 3.0\times 10^{-4}(L_z\Omega)^2$,
$\half \langle \rho v_z^2\rangle_i = 7.0\times 10^{-5}(L_z\Omega)^2$.
The evolution of the ion and neutral kinetic energy is depicted in Fig.
4.  The volume-averaged kinetic energy of the neutrals evolves closely
with that of the ions, except that it is larger by nearly the ratio
$\rho_n/\rho_i$.  The 
ion velocities are slightly larger on average.

As in the single-fluid case, angular momentum is transported outward by
the Reynolds stress (both neutral and ion) and by the Maxwell stress.
The Maxwell term is the largest stress component with $-\langle
B_xB_y/4\pi\rangle = 0.0034 (L_z\Omega)^2$; the neutral and ion Reynolds
stresses are $2.4\times 10^{-3}$ and $3.5\times 10^{-4} (L_z\Omega)^2$
respectively.  The ion Reynolds stress is proportionally larger than
that of the neutrals, once the neutral to ion density ratio is factored
out.  In these units the sum of the stresses is equal to the
Shakura-Sunyaev $\alpha$ value.  Single fluid simulations find that
while the ratio of total stress to the total pressure can vary
substantially from one simulation to another, the ratio of the stress
to the {\it magnetic} pressure is more nearly constant.  The direct
dependence of the stress on the magnetic pressure reflects both the
fundamental role that the magnetic field plays in driving both the
Maxwell and Reynolds stresses, as well as the high degree of
correlation between the radial and toroidal field components.  Clearly
this remains true for ion-neutral systems.  In the fiducial run
$\alpha_{\rm mag}= 0.61$,  a value comparable to that seen in the
single-fluid $z$-field simulations.  Despite its lower relative value
compared with the single-fluid simulation, the radial field remains
highly correlated with the toroidal.  The ratio $\langle B_xB_y
\rangle/\sqrt{\langle B_x^2\rangle\langle B_y^2\rangle}$ is nearly 1
during the linear growth phase and has an average value of 0.7 after
saturation.

The existence of unstable wavenumbers with large values, and the
relatively low growth rate observed in the magnetic field, suggests
that resolution may be influencing the simulation.  Resolution can be
especially important for ion-neutral simulations, because they
typically have a wider range of rapidly growing wavelengths and
physically important lengthscales compared to the single fluid system.
For the parameters of the fiducial run, this is certainly true, and
more grid zones should be required to resolve the linear growth phase
adequately compared to the one-fluid system.  Run LZ1 doubles the
resolution of the fiducial run to $63\times 127\times 63$ grid zones
while maintaining all other initial parameters.  The observed growth
rate for the magnetic field in the high resolution run is $0.35\Omega$
(see Fig. 2), 30\% larger than in the lower resolution run.  The most
obvious mode seen in density plots during the linear growth phase has
$L_z/\lambda = 6$.  Although ultimately both the low and high
resolution simulations are dominated by the largest scales, resolution
remains an important factor throughout the simulation.  The high
resolution simulation has 1.8 times as much magnetic energy as the
fiducial run, when compared over the same time interval after
saturation.  The total stress is larger by a factor of 1.4.

\subsubsection{Effect of ionization fraction } 

In the next set of simulations, we investigate the effect of the
ionization factor by comparing the fiducial run to model Z20 with
$f=0.01$ and $\gamma=0.1$,  and model Z21 with
$f=0.001$ and $\gamma=1$.  We retain the same initial Alfv\'en
speed $v_{A{i}} = 0.044 (L_z\Omega)$ as used in the fiducial run.
Because we hold the Alfv\'en speed constant, the initial $B_z$ magnetic
field decreases as $f^{1/2}$.  Table 2 lists time and space averaged
values after saturation for this set of runs.

Although the parameter $kv_{A{i}}/\Omega = 0.28 L_z/\lambda$ is the
same from one run to the next, the linear growth rates for the
resolvable vertical wavelengths decrease with $f$ (BB).  For $\lambda=
L_z$ the linear growth rates are 0.15, 0.047, and 0.015$\Omega$; for
$\lambda=L_z/2$ the rates are 0.27, 0.093, and 0.03$\Omega$.  The
measured growth rates for $B_x$ in the simulations are 0.27, 0.072, and
0.016$\Omega$ for $f=0.1$, 0.01, and 0.001 respectively.

Despite the reduced growth rates, the total magnetic field {\it
amplification} is comparable in all three runs, about two orders of
magnitude in energy.  The time-averaged magnetic energy after
saturation in these runs is $1.2\times 10^{-2}$, $1.1\times 10^{-3}$,
and $2.0\times 10^{-4} (L_z\Omega)^2$.  These correspond to similar
saturation amplitudes in the ion Alfv\'en speed, $v_{A{i}}^2\sim
0.1\ (L_z\Omega)^2$ (Fig. 5).  The majority of the magnetic energy is
in the toroidal component.  Since the saturated magnetic energy is
proportional to $f$, the total stress is similarly proportional to $f$
as is the Shakura-Sunyaev $\alpha$ parameter, defined with respect to
the total pressure (neutral plus ion pressure).  More significantly,
the relative strength of the radial field compared to the toroidal
declines with $f$.  This means that the average $\alpha_{\rm mag}$ also
declines with $f$; $\alpha_{\rm mag} = 0.6$, 0.2, and 0.1 in these three
runs.

As in the fiducial run, the ions are squeezed into vertically thin
sheets lying between regions of strong toroidal field; the lower the
initial ionization fraction, the greater the compression.  At a
representative late time in runs Z17, Z20, and Z21, the maximum ion
densities are 2.5, 2.6 and 4.2 times the initial value.
The neutrals look similar in all three runs, again
dominated by nearly $z$-independent trailing pressure waves.

Because the three models saturate at comparable ion Alfv\'en speeds, we
conclude that the final state depends most strongly upon the ion
density, that is, the effective inertia for the instability remains
close to $\rho_i$.  At coupling frequencies $\sim 1$ the role of the
neutrals is mainly that of a sink for the
turbulence.  This is consistent with the observation that the magnetic
energies in these simulations are smaller than predicted from equation
(6) for a single-fluid with density equal to $\rho_{i}$.

While these results imply an increasingly small angular momentum
transport for increasingly small ionization factors, this is partly
because we have chosen to initialize this ensemble of simulations with
constant Alfv\'en speed.  An alternative is to begin with constant
magnetic energy density.  For example, a model with $f=0.001$ is still
unstable even with the field is as strong as in the fiducial $f=0.1$
simulation.  With this strength of magnetic field and with $f=0.001$,
we have $\beta_{i} = 0.67$ initially, and $kv_{A{i}}/\Omega =
8.85  L_z/\lambda$.  The analytic growth rate is 0.35 for the $\lambda
= L_z$ mode.  When such a run (Z23) is performed,  however, the
observed growth rate is only $0.15\Omega$ compared with $0.27 \Omega$
in the fiducial run.  The field is sufficiently strong (relative to the
ion pressure), that compressibility is important, and this reduces the
growth rate below the weak-field, incompressible value obtained from
equation (24) of BB.

The saturated magnetic field energy in Z23 is lower than in the
fiducial run, with $B^2/8\pi = 1.3\times 10^{-3} (L_z\Omega)^2$.  This is
sufficiently strong that $\beta_{i} = 0.06$ on average.  The dominance
of the magnetic pressure is again reflected in the narrow, nearly
axisymmetric sheets of ions surrounded by strong toroidal field.  At
the end of the run, the maximum ion density is 8 times the initial
value, the largest
relative compression of any of this series of runs.  The ion sheets
have a similarity of appearance to the channel solutions seen in the
single fluid simulations, except that here field amplification stops
and the growth of the nonaxisymmetric parasitic instability is
suppressed or significantly reduced.  Hence, the thin ion sheets
endure.  Despite  a collisional frequency $\sim \Omega$, the ions and
neutrals are not sufficient well coupled to overcome the complete
dominance of the magnetic field on the ion evolution.  The
perturbed ion velocities are larger than those of the neutrals.  The
neutrals are again modified by the presence of nearly z-independent
trailing pressure waves, except that the amplitude of these waves is
now only $\delta P_n/P_n = 0.01$.

The time-averaged Maxwell stress in Z23 is $4.8\times 10^{-4}
(L_z\Omega)^2$, the ion Reynolds stress is $5.5\times 10^{-7}
(L_z\Omega)^2$, and the neutral Reynolds stress is $7.5\times 10^{-5}
(L_z\Omega)^2$ The stress is dominated by the Maxwell component which
remains proportional to the magnetic energy, with $\alpha_{\rm mag} =
0.4$.  The ion Reynolds stress is proportionally larger than the
neutral stress by a factor of 7, but, due to the low ion fraction, its
absolute value makes no significant contribution.

To summarize, we have examined three decades in ionization fraction
with coupling frequencies near the orbital frequency.  The observed
numerical growth rates are consistent with the linear analysis of BB.
Further, we find that these collision frequencies are insufficient to
involve the neutrals in the evolution beyond providing a drag.  The
saturation amplitudes are determined by the ion Alfv\'en frequency, and
low $\beta_{i}$ values.  Because saturation is determined mainly by the
ions, and magnetic equipartition with the ion thermal energy,
ion-neutral systems with very low ionization fractions and weak
coupling would be expected to saturate at small magnetic field
strengths.

\subsubsection{Effect of the collision frequency} 

The analysis of BB highlights the importance of the ratio of the
collision frequency to the epicyclic frequency.  In the simulations
described so far we have maintained the collision frequency near its
critical value, i.e., comparable to the epicyclic frequency.  We next
investigate the effects of altering the collision frequency by
modifying the drag coefficient $\gamma$.  How does the turbulent
magnetic energy vary as the coupling frequency is increased?  At what
point as the coupling frequency is increased do the neutrals couple
strongly enough with the ions to be significantly affected by the MHD
instability?  At what value is the coupling frequency so low that
the ions essentially decouple from the neutrals?

To investigate these questions, we run simulations 
with the same ionization fraction and initial magnetic field strength,
but with different values of the drag coefficient $\gamma$.  We begin
with the initial conditions of the fiducial run, $f=0.1$ and initial
field strength $v_{Az{i}} = 0.044 L_z\Omega$, and set $\gamma = 0.001$ in
run Z24, $\gamma = 0.1$ in Z25, and $\gamma = 1.0$ in Z28.  Along with
the fiducial run, this produces an ensemble of simulations with
$\gamma\Omega/\rho_i=0.111$, 1.11, 11.1 and 111.1.  Time and space
averaged values from these simulations are given in Table 3.

The evolution of the magnetic energy in these simulations is shown in
Figure 6.  Along with the two-fluid runs, this figure includes two
single-fluid simulations labeled C1 and C2.  Run C1 has $\rho = 0.111$
(the same as $\rho_{i}$ in the fiducial run) and C2 has $\rho=1.111$
(the same as $\rho_{i}+\rho_{n}$ in the fiducial run).  We use the
fiducial run's initial magnetic field strength for both these
simulations, hence C1 and C2 have different initial values of the
Alfv\'en speed, specifically $v_A/\Omega = 0.044$ and $0.139$.  In both
of these control simulations, turbulent saturation occurs after three
orbits, with time-averaged magnetic energies of $0.025 (L_z\Omega)^2$
and $0.083 (L_z\Omega)^2$.  While these values are somewhat below that
predicted from the HGB95 empirical relationship (6), 
the ratio of the turbulent magnetic energies
in these two runs is 0.30,  consistent with the functional dependence
expressed in equation (6).

The $\gamma = 0.001$ run (Z24) is similar to the low density single
fluid calculation (C1).  The ions are again squeezed into the very
narrow filaments between strong magnetic fields while the neutrals are
barely disturbed.  This run is the first case discussed where the ion
kinetic energy exceeds the neutral kinetic energy.  The squeezing of
the ions by the magnetic field is sufficiently great that the ions
sheets become effectively one zone wide and the simulation ends.  An
isothermal single-fluid simulation behaves like Z24.  The C1 run uses
an adiabatic equation of state; the additional pressure keeps the fluid
sheets resolved, permitting the computation to continue.  Again, this
demonstrates that compressibility effects become important when $\beta
< 1$.

The differences between the $\gamma = 0.01$ run (Z17) and the $\gamma =
0.1$ run (Z25) are small but significant.  
The magnetic field energies are comparable,
but Z25 has a 25\% larger average radial magnetic field strength, and
greater neutral Reynolds stress leading to a 20\% larger $\alpha$
value.  Thus, an increase by 10 in the coupling frequency has had some
influence on the saturation level, but has not yet brought the magnetic
energies to near their single fluid values.

When the drag coefficient is increased to $\gamma = 1$ (Z28) the
collisional frequency exceeds the orbital frequency by 111.  The
resulting turbulent magnetic energy is significantly larger than in the
fiducial run, and continues to grow slowly during the simulation.  The
average values between orbits 10 and 14 are $B_x^2/8\pi = 0.0049
(L_z\Omega)^2$, $B_y^2/8\pi=0.050 (L_z\Omega)^2$, and $B_z^2/8\pi =
0.0017 (L_z\Omega)^2$, about 70\% of the values seen in the single
fluid simulation C2.

Table 2 lists the ratios of the ion to neutral kinetic energies.  If
the two fluids were fully coupled, this ratio, and a similar ratio for
the Reynolds stress, would be precisely the ionization fraction.  This
provides one measure of the effective coupling in the turbulent state.
As the coupling frequency increases, the ratio of the
Maxwell stress to neutral Reynolds stress increases toward the
single-fluid range of 4--5.  The ion Reynolds stress also increases, but
even for run Z28 it is not yet quite one tenth the neutral value.  It
appears that coupling frequencies must be $\gamma\Omega/\rho_i \gapprox
100$ before the evolution becomes similar to that of the fully ionized
fluid.

As another test of the role of $\gamma$, we compare a series of
$f=0.01$ runs with $\gamma$ values equal to $0.01$,
$\gamma\Omega/\rho_i=0.1$ (Z19), $\gamma =0.1$, $\gamma\Omega/\rho_i=1$
(Z20), and $\gamma = 10$, $\gamma\Omega/\rho_i=100$ (Z27).  In the
least well-coupled model (Z19) the ions are squeezed into very narrow
filaments between regions of strong toroidal field.  With increasing
$\gamma$ the ions are less confined into coherent sheets, more
turbulent, and more like the single-fluid simulations.  The angular
momentum transport is again dominated by the Maxwell stress.  In the
three runs, $\alpha_{\rm mag} = 0.2$, 0.3, and 0.4.  The increase in
$\alpha_{\rm mag}$ reflects the larger values of radial field that are
obtained with increasing $\gamma$.  In Z27, the ratio of the ion to
neutral kinetic energy and ion to neutral Reynolds stress are nearly
equal to the ionization fraction, indicating good coupling.  In order
of increasing $\gamma$, the turbulent magnetic energies are $2.6 \times
10^{-3}$, $1.0\times 10^{-3}$, and $6.7\times 10^{-3} (L_z\Omega)^2$.
These energies are dominated by the toroidal field; the ratios of
toroidal to radial magnetic energy are 0.006, 0.009, and 0.027.
Although the best-coupled run (Z27) has the largest magnetic energy,
its value is still a factor of 3 below the single-fluid prediction with
density $\rho_{\rm tot}$.  It should be noted that in terms of the
total gas pressure, the initial field is quite weak with $\beta_{\rm
tot} = 10^5$.  For a single fluid simulation with this strength field,
the critical wavelength is less than the grid spacing, so this
simulation is somewhat under-resolved.

Next, we consider a well-coupled model (Z22) with a stronger initial
field and a lower ionization fraction, specifically, $f=0.001$ and
$\gamma = 100$.  This raises the collision frequency to
$\gamma\rho_{i}/\Omega = 100$.  The initial field is set to $\beta_{i}
= 0.1$.  The instability develops rapidly as a channel solution (two
channels in the vertical direction), then saturates at 4.5 orbits as
the channels break up into turbulence.   Despite beginning with a
strong field relative to the ions, the coupling is sufficient to raise
the effective inertia of the system, and to ensure that the neutrals
become fully turbulent as well.  The magnetic energy at the end of the
simulation is $0.036 (L_z\Omega)^2$ which corresponds to $\beta = 28$,
and $\beta_{i} = 0.002$.  As always, the Maxwell stress is the largest
with a value of $0.018 (L_z\Omega)^2$, but the neutral Reynolds stress
is now also significant at $0.0073 (L_z\Omega)^2$.  The ion Reynolds
stress is smaller by a factor of $f$.  The value of $\alpha$ is 0.025,
while $\alpha_{\rm mag}= 0.7$.  Overall the evolution is much like that
for a single fluid.
If the collision frequency is sufficiently large, full MHD turbulence
and significant angular momentum transport are recovered, even for low
ionization fractions.

To explore the other extreme, and to follow the approach to decoupling
in the low-$\gamma$ limit, we run 4 models with $f=0.01$ and
$\gamma=0.1$, 0.05, 0.01, and 0.001 (Z4a, Z5a, Z6a, Z7a).  These models
use an adiabatic equation of state with an initial ratio of ion to
neutral gas pressure of $4\times 10^{-4}$.  The adiabatic equation of
state prevents the magnetic field from squeezing the ions into
numerically unresolvable sheets.  The runs of this group are all very
similar except for Z7a.  Although all four runs end up with similar
toroidal field energies, only Z7a has a substantial radial field
energy; this results in a larger total Maxwell stress.  The average
energy in the radial ion velocity fluctuations is almost two orders of
magnitude larger in Z7a compared with the others.  The observed growth
rate in Z7a is comparable to the single fluid case.  These results
indicate that when $\gamma\rho_i/\Omega \lapprox 0.01$ the ions and
neutrals behave essentially as two distinct fluids.

\subsubsection{Other effects } 

BB determined that in two-fluid systems the presence of a strong
toroidal field can affect growth rates.  We demonstrate this point by
rerunning the fiducial run with the addition of a strong toroidal field
with $B_y = 10B_z$ (YZ1).  Such a field has $v_{A\phi} > 1.6 c_{si}^2$,
and with $\gamma \rho_i/\Omega =1.11$ the growth rates should be
significantly affected (BB).  Indeed, we find that the growth rate of
the perturbed magnetic field drops to $0.18\Omega$.  After
saturation at orbit 7
the radial magnetic and kinetic energies are
reduced compared to the fiducial run, and this reduces the total
transport to $\alpha = 0.002$.  The total magnetic energy is slightly
larger than in the fiducial run, although this is mainly because here
the toroidal field {\it began} at this level.

The results discussed so far point to a significant role for
compressibility in the growth and saturation of the instability.  The
ions are strongly affected by magnetic pressure when $\beta_{i} < 1$
and the ion-neutral coupling frequency is near or below $\sim \Omega$.
To further study the role of compressibility, we rerun the fiducial
case with the ion and neutral masses equal (run Z18); this increases
the ion pressure.  Both this run and the fiducial run are the same
during the linear growth phase.  Run Z18, however, saturates at a
magnetic field energy level that is about twice that of Z17.  Since the
saturated $\beta_{i}$ value is larger in Z18 (even though the magnetic
field is stronger than in Z17), the ions are less tightly confined.
This is consistent with the conclusion that if the saturated magnetic
energies yield $\beta > 1$, the gas pressure remains mostly
unimportant, whereas if $\beta < 1$, pressure can bring field
amplification to an end.  The effects of pressure are particularly
noticeable in the two-fluid simulations because, in contrast to
previously studied single-fluid models, the magnetic pressures in the
saturated states generally have $\beta_{i} < 1$.  (This depends, of
course, on how we chose to initialize the simulations.  Here, the
initial ion pressure is generally much less than the neutral pressure,
and the neutral pressure is chosen so that $c_{sn} = L_z\Omega$.) If the
coupling is sufficiently weak such that the effective inertia for the
magnetic instability is close to that provided by the ions alone, the
pressure prevents the magnetic field from becoming much stronger than
equipartition with the ion pressure.

\subsection{Zero net $B_z$ fields} 

The structure associated with the nonlinear regime of the instability
in the weakly-coupled pure $B_z$ field simulations described above is
dominated by the persistence of the channel solution in the form of
narrow, nearly axisymmetric sheets of ions.  This is in contrast to the 
single-fluid simulations, where parasitic instabilities inevitably cause the
channel solution to break up into MHD turbulence (HGB95).  The
axisymmetric coherence of the channel solution is lost when the
initial $B_z$ field is not uniform.  To study this
in the two-fluid case, we have performed simulations
which begin with fields of the form $B_z \propto \sin (x)$ so that there
is zero net flux through the computational domain.

Our first zero net $B_z$ field simulation (ZN1) is computed using
nearly the same parameter values as the fiducial pure $B_z$ field
simulation, i.e., $\beta_i = 80$, $f=0.1$, and $\gamma= 0.009$,
yielding a coupling frequency of $\gamma \rho_i / \Omega = 1$.  We
adopt an isothermal equation of state, and use the standard resolution
to evolve the model.  Once again, we observe exponential growth of the
magnetic energy, with saturation at an amplitude of $0.01(L_z\Omega)^2$
occurring near 5 orbits.  In the saturated state, the Maxwell stress
dominates the Reynolds stress, with $-\langle B_xB_y/4\pi\rangle =
0.002(L_z\Omega)^2$ averaged over the first 10 orbits.  These values
are all consistent with those reported for the fiducial run Z17.  As is
the case in single-fluid zero net $B_z$ field simulations (Hawley,
Gammie, \& Balbus 1996), the amplitude of the turbulence decreases
significantly after saturation, but it never dies completely away; the
magnetic energy remains at least a factor of 5 larger than in the
initial state.

We have also computed the evolution of a zero net $B_z$ field with a
smaller drag coefficient, i.e. $\gamma= 0.002$ yielding $\gamma \rho_i
/ \Omega = 2/9$, but all other parameters identical to ZN1.  This model
(ZN2) evolves in a similar fashion, but with a smaller initial growth
rate that produces saturation at a slightly later time (6 orbits).
The amplitude of the saturated magnetic energy, kinetic energy, Maxwell
and Reynolds stress are all similar to the pure $B_z$ field simulation
Z24.  However, unlike Z24, the ions do not have a channel-like
structure.

The turbulence observed in these simulations is driven by the
instability acting upon the ions.  Because of drag, turbulent motions
are also driven in the neutrals.  However, the ions and neutrals will
be coupled only on scales greater than $L_{drag} = v_{An}/\gamma \rho_i$.
Since in these models the magnetic field 
saturates at $v_{Ai} \sim c_{si}$ we can write
\begin{equation}
L_{drag} = {\left(f\over 1-f\right)^\half} {\left( m_n\over m_i\right)^\half }
{L_z\Omega\over \gamma \rho_i } = 0.09 {L_z \Omega \over
\gamma\rho_i}.
\end{equation}
For the parameter values adopted
for run ZN1, $L_{drag}=0.09$, and for ZN2 $L_{drag}=0.42$.  To
investigate whether there is any quantifiable difference between the
turbulence in the ions and neutrals on scales above and below
$L_{drag}$, we plot in Figure 7 the power spectrum of the specific kinetic
energy in the fluctuations of the $z$-component of velocity ($\delta
v_z^2$) in
both the ions and neutrals for ZN1 (top panel) and ZN2 (bottom panel).
In run ZN1, the energy in velocity fluctuations in the ions and
neutrals is virtually identical on all scales.  Given the wavenumber
associated with the drag length is $k_{drag} = 22 \pi/L_z$,
which is more than the largest wavenumber representable at the standard
resolution, this result is to be expected.  However, the power spectrum
for run ZN2 shows a systematic difference between the ions and neutrals
above a wavenumber of about 10; the energy associated with
velocity fluctuations in the neutrals is less than that of the
ions by about an order of
magnitude.  In this case, $k_{drag} = 4.8 \pi/L_z$, which is
in agreement with the observed location of the break in the power
spectrum.  In summary, we find that on scales less then $L_{drag}$ the
ions and neutrals are poorly coupled, and the structure of the neutrals
is very smooth, whereas on scales larger than $L_{drag}$ the ions and
neutrals are well coupled and exhibit a decreasing power law spectrum
with identical amplitude.

\subsection{Uniform $B_y$ fields} 

The analysis of BB considered axisymmetric, vertical wavenumber modes.
A remaining question is whether purely toroidal fields are also
unstable in the two-fluid system.  Toroidal fields exhibit a
nonaxisymmetric instability in the single-fluid system, as found both
analytically and numerically (Balbus \& Hawley 1992; HGB95; Matsumoto
\& Tajima 1995).  In the fully ionized disk, the observed growth rates
are lower than with vertical fields, but not dramatically so.  The
turbulent magnetic energy densities that result are lower than those
seen in the vertical field simulations by about an order of magnitude,
although they  are still proportional to the geometric average of the
background field strength and the field strength corresponding to the
largest possible unstable wavelengths given the dimensions of the
simulation box, $L_y$ (HGB95).  Outward angular momentum transport is
produced with a total stress that remains proportional to the magnetic
pressure, with $\alpha_{\rm mag} \sim 0.5$.

The nonaxisymmetric instability must be present for the two-fluid
system as well, at least in completely coupled and decoupled
limits.  Here we investigate what happens near the critical value of
the collision frequency, $\gamma \rho_i/\Omega = 1$ with the series of
toroidal field runs listed in Table 4.  Figure 8 shows the time
evolution of the toroidal and radial magnetic field energies for the
models listed in Table 4.  Table 5 lists the value of some averaged
quantities after saturation for these runs.

The first simulation (Y5) has an ion fraction of $f=0.1$ and a drag
coefficient $\gamma = 0.01$.  Again we use an isothermal equation of
state, and assume a ratio of ion to neutral mass $m_{i}/m_{n}$ of
$30/2.33$.  The initial field is a uniform $B_y$ field with strength
$\beta_{i} = 10$, which corresponds to a value of $kv_{A{
i}}/\Omega = 0.125 L_y/\lambda$.  The perturbations grow slowly in this
simulation, leveling out after 20 orbits, continuing to grow very
slowly after that time.  Relatively little total field amplification
occurs, with the average magnetic field strength near the end of the
run (46 orbits) equal to $2.1\times 10^{-3}$.  Total stress is also
small, with $\alpha = 4.4\times 10^{-4}$ and $\alpha_{\rm
mag} = 0.22$.  The ions are found in small scale features that have a
mostly $m=1$ structure.  The perturbation in the neutrals consists of
trailing $m=1$ waves that are nearly independent of height $z$.

In these simulations, the fastest growing unstable wavelengths are
short compared to the computational domain size, although the full
range of unstable wavelengths is large and includes lengths comparable
to $L_y$.  Because of this, numerical resolution will play an important
role in the linear growth and possibly in the final saturation of the
instability.  To measure this effect, we repeat the fiducial toroidal
field run with twice the resolution, i.e., $63\times 127\times 63$
(LY1).  The observed growth rate during the initial orbits is
increased.  At the end of the high resolution run the magnetic energy
is twice the value at the same time in Y5.  At this point the growth
rate levels off to a value much like the lower resolution simulation.
Given the low growth rates it seems impractical to carry out this
simulation to late times, but it is clear that resolution significantly
affects the initial evolution of these toroidal field models.  In
analyzing these runs, therefore, we will focus on qualitative effects
and relative values of quantities.

In the next model (Y6) we increase the initial $B_y$ field to
$\beta_{i} = 0.8$, which corresponds to $kv_{A{i}}/\Omega = 0.441
L_y/\lambda$.  This is the same ion Alfv\'en speed as in the toroidal
plus $B_z$ run discussed above (YZ1).  The initial growth rate is
larger than in Y5, but at saturation, which occurs at orbit 15, the
radial field energy has risen only to $1.5\times 10^{-4}
(L_z\Omega)^2$.  Again the stress levels are low, with $\alpha =
8.4\times 10^{-4}$ and $\alpha_{\rm mag} = 0.072$.  The low stress
levels, and, in particular, the low level of $\alpha_{\rm mag}$ is due
to the weak amplification of the poloidal field, and its small value
compared to the initial toroidal field.

Generally, larger growth rates are obtained when $kv_A/\Omega\sim 1$,
but with Y6 we are already in the regime where the initial field is
superthermal (in the ions).  The linear analysis of BB and the single
fluid analysis of Balbus \& Hawley (1992) found that compressibility
becomes important for such strong fields, reducing the effective growth
rate.  In the present simulations the ion pressure can be increased and
the value of $\beta_{i}$ decreased for a given field strength if
the ion and neutral masses are set equal, $m_{i}=m_{n}$.  We
repeat the above two runs with this change, carrying out runs with
$v_{Az{i}}/\Omega = 0.441$ (Y7; compare with Y6) and $v_{Az{
i}}/\Omega = 0.125$ (Y8; compare with Y5).  The increase in ion pressure
has relatively little effect on the evolution during the linear phase.
After 20 orbits, however, the field in run Y8 grows to higher levels,
suggesting that the ion pressure plays a significant role in the
nonlinear evolution.  This is consistent with the vertical field
results discussed above.  In contrast, comparing runs Y6 and Y7 (with
$\beta_{i} =  0.8$ and 10.3 respectively) produces a noticeable
difference during the linear phase.  The growth rate is increased and
the total field is amplified by a factor of 2.3 to $\beta_{i} =
4.4$.  Angular momentum transport is similarly enhanced, with
$\alpha=0.008$ and $\alpha_{\rm mag} = 0.33$.  These runs confirm
that compressibility can affect both the linear and nonlinear
evolution of the instability whenever $\beta_{i} < 1$.

As the collision frequency increases, the system should approach that
of a single fluid.  To test this we repeat the fiducial run with the
collision parameter increased to $\gamma \rho_{i}/\Omega = 111$
(Y9).  In this run the initial growth through 15 orbits is quite
similar to the first run (Y5), but poloidal field growth continues for
a longer time than in Y5.  At 40 orbits the field energies are still
rising, and have twice the magnetic energy as Y5.  The tight coupling
means that the perturbed velocities in the two fluids have the same
values on average.

\section{Discussion}

We have carried out a series of simulations to examine the behavior of
the magnetorotational instability in partially-coupled ion-neutral
systems.  We find that weak vertical fields are unstable with growth
rates consistent with the linear analysis of BB.  The initial evolution
of the vertical field instability behaves much like the channel
solution found in the single fluid simulations.  This is consistent
with the finding of BB that the unstable two-fluid mode with vertical
magnetic field is still an exact local nonlinear solution to the
equations of motion.  In our simulations, if the ion fraction and the
collision frequency are low the channel solution can persist, squeezing
the ions into narrow channels between regions of strong toroidal
field.  As $\beta_i$ decreases, the simulation ends when
the channels become one grid zone thick.

Brandenburg \& Zweibel (1995) have emphasized that because ion-neutral
coupling acts as a nonlinear diffusion process, certain initial field
configurations can evolve into thin current sheets.  Because we are far
from the well-coupled limit studied by Brandeburg \& Zweibel, it is
difficult to asses the relative importance of nonlinear diffusion due
to ambipolar diffusion in these simulations.  We have identified the
primary mechanism for the growth of the narrow sheets in the vertical
field simulations with the ``channel solution'' of the
magnetorotational instability.  Precisely the same thin sheets are
observed in single-fluid (fully ionized) simulations, and in the ions
in the nearly-decoupled limit.  In both cases, the fluid is squeezed by
an exponentially growing field until the channel becomes one zone thick
and the simulation ends.  Further, the zero-net-field two-fluid
simulations presented here, which do not produce a channel solution,
also avoid the formation of large-scale current sheets.  It remains an
open question whether the nonlinear diffusion of Brandenburg \& Zweibel
(1995) increases the tendency to thin sheets, or otherwise affects the
turbulent state in ion-neutral systems.

Through a series of simulations with varying drag coefficient $\gamma$
we have explored the transition from the fully coupled to the
completely uncoupled limit.  We find that there are important two-fluid
effects over the range $0.01 < \gamma\rho_{i}/\Omega < 100$.  In
simulations at the upper end of this range the ions and neutrals are
strongly coupled, although the turbulence levels remain somewhat below
the single-fluid case.  Our results are consistent with the simulations
of Mac Low et al. (1995) who found significant ambipolar diffusion when
$\gamma\rho_i/\Omega =30$, but little apparent diffusion at a value of
300.

At the other extreme, the weakly coupled limit with
$\gamma\rho_{i}/\Omega \sim 0.01$, the perturbed ion kinetic energy
actually exceeds that of the neutrals (despite an ionization fraction
of 0.01), and the magnetic energy is increased over levels found in the
results for larger drag coefficient values.  This signals the almost complete
separation of the ion and neutral fluid.  

When the collisional frequency is comparable to the orbital frequency,
$\gamma\rho_{i}/\Omega \sim 1$, the nonlinear evolution of the
instability is primarily controlled by the ion density.  Magnetic field
saturation occurs near or just above equipartition with the ion
pressure.  Hence the saturated magnetic field energy goes down with
ionization fraction,  which, in turn, means that the total angular
momentum transport is also proportional to $f$.

The instability transports angular momentum in the two-fluid system,
through a combination of Maxwell and Reynolds stresses.  The total
stress is proportional to the magnetic energy density in all
simulations.  The constant of proportionality ($\alpha_{\rm mag}$)
depends on the amount of radial field amplification relative to the
toroidal field.  With good coupling or larger ionization fractions
$\alpha_{\rm mag}$ is comparable with the single-fluid value (HGB95).
Weak coupling (but not completely uncoupled) and low ionization
fractions produce reduced values.

Because the neutrals are coupled to the magnetic field only through
drag with the ions, the effect of the instability on the neutrals
depends on the size of the drag coefficient.  The neutral system alone
is hydrodynamic, and as such is stable to hydrodynamic perturbations
(BHS96).  The ion velocity fluctuations are correlated so as to produce
a net positive ion Reynolds stress, and that correlation is conveyed to
the neutrals.  However, because the neutrals lack direct coupling to the
magnetic fields, a positive Reynolds stress acts as a sink in the
neutral angular velocity fluctuation equation.  In the intermediate
coupling regime this apparently works to damp the strength of the ion
turbulence to levels below what would be seen with the ion system fully
decoupled.  

Why does gas pressure matter in these simulations and not in the
single-fluid studies done  previously?  In the single-fluid simulations
(HGB95) the gas pressure was largely irrelevant,  in part because the
box size was chosen so that $P=(L_z\Omega)^2$ and the largest unstable
wavelength corresponded to a subthermal field.  In the ion-neutral
system as we have set it up here, fields that are superthermal in the
ion pressure (i.e., have $\beta_{i} < 1$) can still be unstable.  For
small ionization fractions these fields will be very subthermal in
terms of the total (neutral plus ion) pressure ($\beta \gg 1$).  When
$\gamma\Omega/\rho_i \sim 1$, the ion pressure significantly affects
the nonlinear stage of the evolution, for both toroidal and vertical
initial field models.  The tendency for the ions to form thin sheets
between regions of strong toroidal field is symptomatic.  In some cases
this reduces the vigor of the resulting turbulence.  Stronger
turbulence and more angular momentum transport can be obtained by
decreasing the ion mass, and hence increasing the ion pressure (under
the assumption of equipartition of thermal energy between the ion and
neutral particles, $m_{i}c_{si}^2 = m_{n} c_{ni}^2$).  We consistently
find stronger turbulence with higher ion pressures.

When toroidal fields are included in the initial conditions along with
vertical field, they can reduce linear growth rates, in accordance with
the analysis of BB.  However, the reduced growth rates have less of an
effect once the nonlinear regime is reached.  In fact, far from
preventing the the growth of the instability, we find that purely
toroidal initial fields are also unstable.  However, the instability is
significantly reduced for weak coupling
if the toroidal field is superthermal.  Although this is true in
the single fluid case as well, there it was hardly an issue because
$v_A \sim c_s$ was already a very strong field.  Here, however, if
$m_{i} > m_{n}$, the ion pressure can be substantially smaller
than the neutral.  If the ion fraction is small, the ion Alfv\'en speed
can easily become large compared to the ion sound speed, despite a
relatively weak toroidal field.

Numerical resolution is more important in these simulations than for
the single fluid models.  The presence of both ion and neutral
components introduces additional lengthscales that can be quite
disparate, as in the case of low ionization fractions and intermediate
strength coupling.  A Fourier analysis of the velocity fluctuations
associated with the ions and neutrals shows that, as expected, on
scales greater than $L_{drag} = v_{An}/\gamma \rho_i$, the ions
and neutrals are well coupled with identical power spectra for
velocity fluctuation, whereas on scales less than $L_{drag}$ the
fluctuations in the neutrals are at a considerably lower amplitude
than those in the ions.  Thus, $L_{drag}$ represents an important
lengthscale introduced into two fluid models that must be resolved
numerically.

In summary, we expect that the two-fluid effects studied here will be
appropriate to a transition region in protostellar or protoplanetary
accretion disks between the inner, hot, and fully ionized regions, and
the outer, cold, and essentially neutral regions.  Our simulations
suggest a more stringent criterion for good coupling than that obtained
from the linear result of BB.  We find that, while the linear
instability is present for coupling frequencies $\sim \Omega$,
significant turbulence and angular momentum transport can only occur
when coupling frequencies $\gamma\Omega/\rho_i > 100$.  For weaker
coupling, the magnetic energies are determined primarily by the ion
density.

There are many interesting issues regarding this transition region in a
protostellar disk that go beyond the limitations of this initial
study.  Here we have assumed that both the recombination and ionization
timescales are much longer than the orbital period.  In real systems,
the ion density in structures such as the ion filaments may be limited
by recombination.  On the other hand, at the interface between weak and
strong coupling, increasing the ionization fraction increases the
efficiency of magnetic coupling, which in turn increases the level of
turbulent heating thus raising the ionizational level yet higher.
Investigating stability issues such as these, as well as the global
structure and dynamics of weakly coupled disks, are fruitful areas for
future research.

We thank Steven Balbus for valuable discussions,
and Omer Blaes and Mordecai-Mark Mac Low for comments on the
manuscript.   This work is supported in part by NASA grants NAG-53058,
NAGW-4431, and NSF grant AST-9423187 to J.H., by NASA grant NAG-54278
and NSF grant AST-9528299 to J.S., and by a metacenter grant MCA95C003
from the Pittsburgh Supercomputing Center and NCSA.

\begin{table}
\centerline{\bf TABLE 1: Z FIELD SIMULATIONS}
\vspace{3mm}
\begin{center}
\begin{tabular}{ccccccccc}
\tableline
Model &$f$&$\gamma$&$\gamma\rho_i/\Omega$&$m_i/m_n$&$v_{Azi}/L_z\Omega$
&Grid &Orbits&Comment\\
\tableline
Z17&  0.1& 0.01& 1.1&12.9&0.044& $31\times 63 \times 31$& 26&Fiducial \\
LZ1&  0.1& 0.01& 1.1&12.9&0.044& $63\times 127\times 63$& 13&High-Res\\
Z20& 0.01&  0.1& 1.0&12.9&0.044& $31\times 63 \times 31$& 30\\
Z21&0.001&    1& 1.0&12.9&0.044& $31\times 63 \times 31$& 71\\
Z23&0.001&    1& 1.0&12.9& 0.46& $31\times 63 \times 31$& 10\\
Z24&  0.1&0.001&0.11&12.9&0.044& $31\times 63 \times 31$& 4\\
Z25&  0.1&  0.1&11.1&12.9&0.044& $31\times 63 \times 31$& 11\\
Z28&  0.1&    1& 111&12.9&0.044& $31\times 63 \times 31$& 14\\
Z19& 0.01&  0.1& 0.1&12.9&0.044& $31\times 63 \times 31$& 27\\
Z27& 0.01&  10.& 101&12.9&0.044& $31\times 63 \times 31$& 29\\
Z22&0.001&  100& 100&12.9& 1.25& $31\times 63 \times 31$& 63\\
Z4a& 0.01&  0.1& 1.0& 2.1&0.044& $31\times 63 \times 31$& 32&Adiabatic\\
Z5a& 0.01& 0.05& 0.5& 2.1&0.044& $31\times 63 \times 31$& 32&Adiabatic\\
Z6a& 0.01& 0.01& 0.1& 2.1&0.044& $31\times 63 \times 31$& 24&Adiabatic\\
Z7a& 0.01&0.001&0.01& 2.1&0.044& $31\times 63 \times 31$& 19&Adiabatic\\
YZ1&  0.1&    1& 1.1&12.9&0.044& $31\times 63 \times 31$& 14\\
Z18&  0.1& 0.01& 1.1&   1&0.044& $31\times 63 \times 31$& 15\\
\tableline
\end{tabular}
\end{center}
\end{table}

\begin{table}
\centerline{\bf TABLE 2: TIME- AND VOLUME-AVERAGE VALUES}
\centerline{\bf EFFECT OF IONIZATION FRACTION}
\vspace{3mm}
\setlength{\arraycolsep}{0.5cm}
\begin{center}
\begin{tabular}{ccccc}
\tableline
$\langle\langle{\rm Quantity}\rangle\rangle$& Z17& Z20& Z21& Z23 \\
\tableline
$f$&0.1&0.01&0.001&0.001\\
${B_x^2 /{8\pi}(L_z\Omega)^2}$
&$5.3\times 10^{-4}$
&$7.8\times 10^{-6}$
&$1.8\times 10^{-7}$
&$5.7\times 10^{-5}$\\
${B_y^2 /{8\pi}(L_z\Omega)^2}$
&$1.1\times 10^{-2}$
&$1.1\times 10^{-3}$
&$2.0\times 10^{-4}$
&$1.1\times 10^{-3}$ \\
${B_z^2 /{8\pi}(L_z\Omega)^2}$
&$3.8\times 10^{-4}$
&$1.4\times 10^{-5}$
&$1.2\times 10^{-6}$
&$1.2\times 10^{-4}$\\
${(\rho v_x^2)_{\rm i} /2(L_z\Omega)^2}$
&$6.8\times 10^{-4}$
&$4.7\times 10^{-6}$
&$7.0\times 10^{-7}$
&$8.1\times 10^{-7}$ \\
${(\rho\delta v_y^2)_{\rm i}/2(L_z\Omega)^2}$
&$3.0\times 10^{-4}$
&$5.5\times 10^{-7}$
&$7.7\times 10^{-8}$
&$1.4\times 10^{-7}$\\
${(\rho v_z^2)_{\rm i} /2(L_z\Omega)^2}$
&$7.0\times 10^{-5}$
&$3.2\times 10^{-7}$
&$8.1\times 10^{-8}$
&$1.3\times 10^{-7}$\\
${(\rho v_x^2)_{\rm i}/(\rho v_x^2)_{\rm n}}$
&0.10 & 0.010 & 0.0010 & 0.0031 \\
${(\rho\delta v_y^2)_{\rm i}/(\rho\delta v_y^2)_{\rm n}}$
& 0.16 & 0.013 & 0.0010 & 0.0048 \\
${(\rho v_z^2)_{\rm i} /(\rho v_z^2)_{\rm n}}$
& 0.13 & 0.013 & 0.0017 & 0.0026 \\
${E_M/KE_{\rm i}}$& 11.8 & 200 & 5100 & 1200 \\
${-B_x B_y /{4\pi}(L_z\Omega)^2}$
&$3.4\times 10^{-3}$
&$1.4\times 10^{-4}$
&$9.0\times 10^{-6}$
&$4.8\times 10^{-4}$\\
${(\rho v_x v_y)_{\rm i}/(L_z\Omega)^2}$
&$3.5\times 10^{-4}$
&$1.4\times 10^{-6}$
&$1.8\times 10^{-8}$
&$5.5\times 10^{-7}$\\
${(\rho v_x v_y)_{\rm n}/(L_z\Omega)^2}$
&$2.4\times 10^{-3}$
&$1.2\times 10^{-4}$
&$1.7\times 10^{-5}$
&$7.5\times 10^{-5}$\\
${\rm Max/Reynolds_i}$\tablenotemark{a}
&9.7& 100 & 510 & 870 \\
${\rm Max/Reynolds_n}$&1.4 & 1.3 & 0.53 & 6.4 \\
${\alpha}$&$6.1\times10^{-3}$
&$2.6\times 10^{-4}$
&$2.6\times 10^{-5}$
&$5.6\times 10^{-4}$\\
${\alpha_{\rm mag}}$ & 0.61 & 0.23 & 0.13 & 0.42 \\
\tableline
\end{tabular}
\tablenotetext{a}{${\rm Max/Reynolds_i}$ is the ratio of the Maxwell 
stress to ion Reynolds stress.  Similarly ${\rm Max/Reynolds_n}$ is 
the ratio of the Maxwell stress to neutral Reynolds stress.}
\end{center}
\end{table}

\begin{table}
\centerline{\bf TABLE 3:  TIME- AND VOLUME-AVERAGE VALUES}
\centerline{\bf EFFECT OF COUPLING CONSTANT}
\vspace{3mm}
\setlength{\arraycolsep}{0.5cm}
\begin{center}
\begin{tabular}{ccccc}
\tableline
$\langle\langle{\rm Quantity}\rangle\rangle$& Z24& Z17& Z25& Z28 \\
\tableline
$\gamma$&0.001&0.01&0.1&1.0\\
${B_x^2 /{8\pi}(L_z\Omega)^2}$
&$2.4\times 10^{-3}$
&$5.3\times 10^{-4}$
&$8.1\times 10^{-4}$
&$4.9\times 10^{-3}$\\
${B_y^2 /{8\pi}(L_z\Omega)^2}$
&$1.9\times 10^{-2}$
&$1.1\times 10^{-2}$
&$1.1\times 10^{-2}$
&$5.0\times 10^{-2}$ \\
${B_z^2 /{8\pi}(L_z\Omega)^2}$
&$3.7\times 10^{-4}$
&$3.8\times 10^{-4}$
&$3.7\times 10^{-4}$
&$1.7\times 10^{-3}$\\
${(\rho v_x^2)_{\rm i} /2(L_z\Omega)^2}$
&$1.2\times 10^{-3}$
&$6.8\times 10^{-4}$
&$5.1\times 10^{-4}$
&$1.3\times 10^{-3}$\\
${(\rho\delta v_y^2)_{\rm i}/2(L_z\Omega)^2}$
&$1.7\times 10^{-3}$
&$3.0\times 10^{-4}$
&$1.4\times 10^{-4}$
&$5.8\times 10^{-4}$\\
${(\rho v_z^2)_{\rm i} /2(L_z\Omega)^2}$
&$2.3\times 10^{-4}$
&$7.0\times 10^{-5}$
&$8.7\times 10^{-5}$
&$4.1\times 10^{-4}$\\
${(\rho v_x^2)_{\rm i}/(\rho v_x^2)_{\rm n}}$
& 1.04&0.10 & 0.11 & 0.11 \\
${(\rho \delta v_y^2)_{\rm i}/(\rho \delta v_y^2)_{\rm n}}$
&5.4  & 0.16 & 0.13 & 0.12 \\
${(\rho v_z^2)_{\rm i} /(\rho v_z^2)_{\rm n}}$
& 0.65& 0.13 & 0.12 & 0.11 \\
${E_M/KE_{\rm i}}$&6.1 & 11.8 & 18 & 26 \\
${-B_x B_y /{4\pi}(L_z\Omega)^2}$
&$1.2\times 10^{-2}$
&$3.4\times 10^{-3}$
&$4.7\times 10^{-3}$
&$2.4\times 10^{-2}$\\
${(\rho v_x v_y)_{\rm i}/(L_z\Omega)^2}$
&$2.2\times 10^{-3}$
&$3.5\times 10^{-4}$
&$2.3\times 10^{-4}$
&$6.4\times 10^{-4}$ \\
${(\rho v_x v_y)_{\rm n}/(L_z\Omega)^2}$
&$7.6\times 10^{-5}$
&$2.4\times 10^{-3}$
&$1.8\times 10^{-3}$
&$5.5\times 10^{-3}$\\
${\rm Max/Reynolds_i}$&5.2 &9.7& 20 & 37 \\
${\rm Max/Reynolds_n}$&150 &1.4 & 2.6 & 4.3  \\
${\alpha}$
&$1.4\times 10^{-2}$
&$6.1\times10^{-3}$
&$6.6\times 10^{-3}$
&$3.0\times 10^{-2}$\\
${\alpha_{\rm mag}}$&0.63 & 0.61 & 0.54 & 0.53  \\
\tableline
\end{tabular}
\end{center}
\end{table}

\begin{table}[p]
\centerline{\bf TABLE 4:  Y FIELD SIMULATIONS}
\vspace{3mm}
\setlength{\arraycolsep}{0.5cm}
\begin{center}
\begin{tabular}{ccccccccc}
\tableline
Model &$f$&$\gamma$&$\gamma\rho_i/\Omega$&$m_i/m_n$&$v_{Azi}/L_z\Omega$
&Grid &Orbits\\
\tableline
Y5 &  0.1&0.01&$ 1.1$&$12.9$&0.125&$31\times 63 \times 31$&46\\
LY1&  0.1&0.01&$ 1.1$&$12.9$&0.125&$63\times 127\times 63$&19\\
Y6 &  0.1&0.01&$ 1.1$&$12.9$&0.441&$31\times 63 \times 31$&32\\
Y7 &  0.1&0.01&$ 1.1$&$   1$&0.441&$31\times 63 \times 31$&17\\
Y8 &  0.1&0.01&$ 1.1$&$   1$&0.125&$31\times 63 \times 31$&50\\
Y9 &  0.1&   1&$ 111$&$12.9$&0.125&$31\times 63 \times 31$&42\\
\tableline
\end{tabular}
\end{center}
\end{table}

\begin{table}[p]
\centerline{\bf TABLE 5:  TIME- AND VOLUME-AVERAGE VALUES}
\centerline{\bf Y FIELD SIMULATIONS}
\vspace{3mm}
\setlength{\arraycolsep}{0.5cm}
\begin{center}
\begin{tabular}{cccccc}
\tableline
$\langle\langle{\rm Quantity}\rangle\rangle$& Y5& Y6& Y7& Y8 &Y9 \\
\tableline
${B_x^2 /{8\pi}(L_z\Omega)^2}$
&$1.3\times 10^{-5}$
&$1.5\times 10^{-4}$
&$1.1\times 10^{-3}$
&$7.1\times 10^{-5}$
&$2.1\times 10^{-4}$\\
${B_y^2 /{8\pi}(L_z\Omega)^2}$
&$2.1\times 10^{-3}$
&$1.2\times 10^{-2}$
&$2.4\times 10^{-2}$
&$4.2\times 10^{-3}$
&$5.6\times 10^{-3}$\\
${B_z^2 /{8\pi}(L_z\Omega)^2}$
&$2.6\times 10^{-6}$
&$8.9\times 10^{-5}$
&$4.0\times 10^{-4}$
&$1.7\times 10^{-5}$
&$6.8\times 10^{-5}$\\
${(\rho v_x^2)_{\rm i}/2(L_z\Omega)^2}$
&$5.9\times 10^{-5}$
&$5.9\times 10^{-5}$
&$5.6\times 10^{-4}$
&$1.7\times 10^{-4}$
&$1.5\times 10^{-4}$\\
${(\rho\delta v_y^2)_{\rm i}/2(L_z\Omega)^2}$
&$1.6\times 10^{-6}$
&$3.8\times 10^{-5}$
&$3.3\times 10^{-4}$
&$3.7\times 10^{-5}$
&$4.1\times 10^{-5}$\\
${(\rho v_z^2)_{\rm i}/2(L_z\Omega)^2}$
&$1.7\times 10^{-6}$
&$1.7\times 10^{-5}$
&$1.1\times 10^{-4}$
&$1.2\times 10^{-5}$
&$3.1\times 10^{-5}$\\
${(\rho v_x^2)_{\rm i}/(\rho v_x^2)_{\rm n}}$
&0.094 &0.097& 0.10 & 0.11  & 0.11\\
${(\rho \delta v_y^2)_{\rm i}/(\rho \delta v_y^2)_{\rm n}}$
&0.11 & 0.19 & 0.21 & 0.15 &0.11\\
${(\rho v_z^2)_{\rm i} /(\rho v_z^2)_{\rm n}}$
& 0.12& 0.10 & 0.096& 0.11 & 0.11 \\
${E_M/KE_{\rm i}}$&35 & 113. & 27.& 25 & 27.\\
${-B_x B_y /{4\pi}(L_z\Omega)^2}$
&$1.8\times 10^{-4}$ 
&$5.6\times 10^{-4}$ 
&$5.7\times 10^{-3}$
&$7.6\times 10^{-4}$ 
&$1.4\times 10^{-3}$\\
${(\rho v_x v_y)_{\rm i}/(L_z\Omega)^2}$
&$2.4\times 10^{-5}$
&$3.4\times 10^{-5}$
&$3.3\times 10^{-4}$
&$5.8\times 10^{-5}$
&$5.5\times 10^{-4}$ \\
${(\rho v_x v_y)_{\rm n}/(L_z\Omega)^2}$
&$2.4\times 10^{-4}$
&$2.5\times 10^{-4}$
&$2.4\times 10^{-3}$
&$4.5\times 10^{-4}$
&$5.0\times 10^{-4}$\\
${\rm Max/Reynolds_i}$&7.6 &17.& 17. & 13. & 26.\\
${\rm Max/Reynolds_n}$&0.75 &2.3 & 2.3 & 1.7 & 2.9 \\
${\alpha}$
&$4.4\times 10^{-4} $
&$8.4\times10^{-4} $
&$7.6\times 10^{-3}$
&$1.1\times 10^{-3} $
&$2.0\times 10^{-3}$ \\
${\alpha_{\rm mag}}$&0.22 & 0.07 & 0.33 & 0.30 &0.34  \\
\tableline
\end{tabular}
\end{center}
\end{table}

\clearpage
\begin{center}
{\bf References}
\end{center}

\refindent Balbus, S.~A., \& Hawley, J.~F. 1991, ApJ, 376, 214
\refindent Balbus, S.~A., \& Hawley, J.~F. 1992, ApJ, 400, 610
\refindent Balbus, S.A., Hawley, J.F., \& Stone, J.M. 1996, ApJ, 467,
76 (BHS96)
\refindent Blaes, O.~M., \& Balbus, S.~A. 1994, ApJ, 421, 163 (BB)
\refindent Brandenburg, A., Nordlund, \AA, Stein, R.~F., Torkelsson,
U. 1995, ApJ, 446, 741
\refindent Brandenburg, A., \& Zweibel, E. G. 1995, ApJ, 448, 734
\refindent Cabot, W. 1996, ApJ, 465, 874
\refindent Cuzzi, J.~N., Dobrovolskis, A.~R., \& Hogan, R.~C. 1996, in
Chondrules and the Protoplanetary Disk, ed. R. H. Hewins, R. H.
Jones, \& E. R. D. Scott (Cambridge: Cambridge Univ. Press), 35
\refindent Draine, B.~T., Roberge, W.~G., \& Dalgarno, A. 1983, ApJ,
264, 485
\refindent Goodman J. \& Xu, G. 1994, ApJ, 432, 213
\refindent Hawley, J.~F., \& Balbus, S.~A. 1991, ApJ, 376, 223
\refindent Hawley, J.~F., \& Balbus, S.~A. 1992, ApJ, 400, 595
\refindent Hawley, J.~F., Gammie, C.~F., Balbus, S.~A. 1995, ApJ, 440,
742 (HGB95)
\refindent Hawley, J.~F., Gammie, C.~F., Balbus, S.~A. 1996, ApJ, 464, 690
\refindent Hawley, J.~F. \& Stone, J.~M.  1995, Comput Phys Comm, 89, 127
\refindent Hill, G.~W. 1878. Amer. J. Math., 1, 5
\refindent Mac Low, M.-M., Norman, M.~L., K\"onigl, A., \& Wardle, M.
1995, ApJ, 442
\refindent Matsumoto, R. \& Tajima, T 1995, ApJ, 445, 767
\refindent Ryu, D. \& Goodman, J. 1994, ApJ, 422, 269
\refindent Shakura, N.~I., \& Sunyaev, R.~A. 1973, A\&A, 24, 337
\refindent Stone, J.~M. 1997, ApJ, 487, 271
\refindent Stone, J.~M., \& Balbus, S.~A. 1996, ApJ, 464, 364
\refindent Stone, J.~M., \& Norman, M.~L. 1992a, ApJS, 80, 753
\refindent Stone, J.~M., \& Norman, M.~L. 1992b, ApJS, 80, 791

\newpage

\begin{figure}
\caption {Numerical growth rates for the radial field in
two-dimensional simulations with ionization fraction $f=0.1$.  The
solid squares are the $64\times 64$ grid zone simulations (5 cases);
the open stars are the $32\times 32$ grid zone simulations (three
cases).  The solid line is the analytic growth rate from linear
stability theory.}
\end{figure}

\begin{figure}
\caption{ Time evolution of the individual components of the
magnetic energy for the fiducial run
Z17 (bold lines) and the high resolution version of the fiducial run
(LZ1). }
\end{figure}

\begin{figure}
\caption{ Volumetric rendering of (a) the ion density and (b) the
magnitude of the toroidal field in the fiducial run at orbit 15.
In (a) brightness is a function of density, whereas in (b) the dark
regions correspond to strong field.  Comparing the two figures shows
that the ions lie in thin sheets sandwiched between regions of strong field.}
\end{figure}

\begin{figure}
\caption{ Time evolution of the ion and neutral perturbed kinetic energies
in the fiducial run (Z17).  The behavior of the two curves is very
similar; they are offset by the neutral/ion density ratio.}
\end{figure}

\begin{figure}
\caption{ Time evolution of the magnetic energy, normalized as the
Alfv\'en speed squared for runs Z17, Z20 and Z21 which have ionization
fractions of 0.1, 0.01, and 0.001, respectively.  While the total
toroidal field amplification is comparable in the three runs, lower
ionization fractions produce smaller poloidal field amplification.}
\end{figure}

\begin{figure}
\caption{  Time evolution of total magnetic energy for a series of
runs with increasing drag coefficient $\gamma$.  The curves are
labeled by their run number, as listed in Table 1.  Also included are
two single-fluid comparison runs, C1 (with density corresponding to the
ion density in the fiducial run) and C2 (with density corresponding to
the total density in the fiducial run).}
\end{figure}

\begin{figure}
\caption{ Power in fluctuations in $v_{zi}$ (solid) and $v_{zn}$
(dashed) versus wavenumber in the y-direction for the zero net field
runs ZN1 (top panel) and ZN2 (bottom panel).  The ``drag length" is
given by eq. (7); $L_{drag}=0.09L_z$ for the top panel, and
$L_{drag}=0.42L_z$ for the bottom.  In ZN2 there is more power at high
wavenumbers in the ions than in the neutrals, while in ZN1 they are both
comparable.  This agrees with the expectation  that when the drag
length is large (bottom), the ions and neutrals are only weakly
coupled, and the ions should show more small scale structure.}
\end{figure}

\begin{figure}
\caption{ Time evolution of toroidal (top) and radial (bottom)
magnetic field energies in the initial toroidal field runs.  The curves
are labeled by the run number as listed in Table 4.  All show field
amplification, although at lower growth rates than the vertical field
models.}
\end{figure}

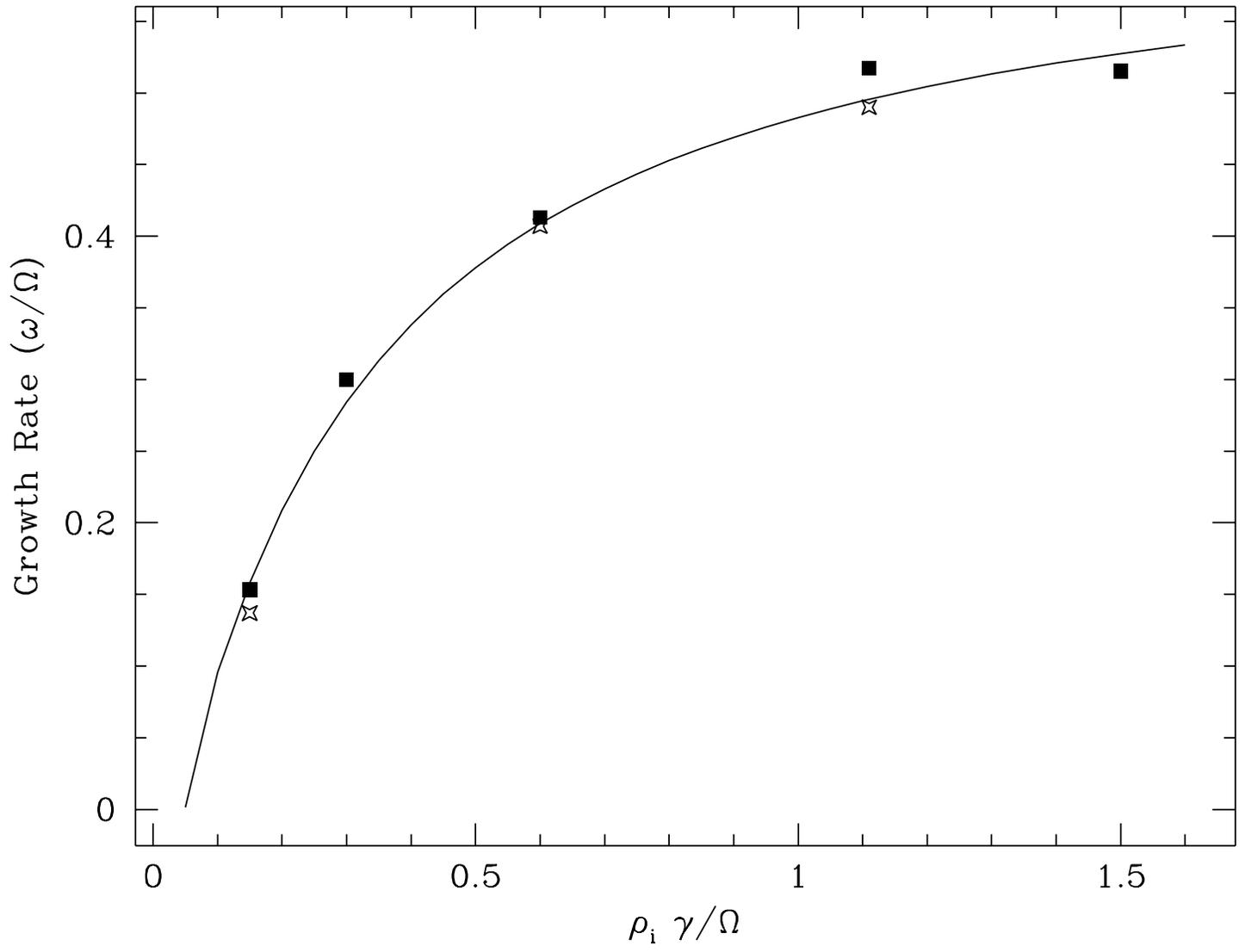
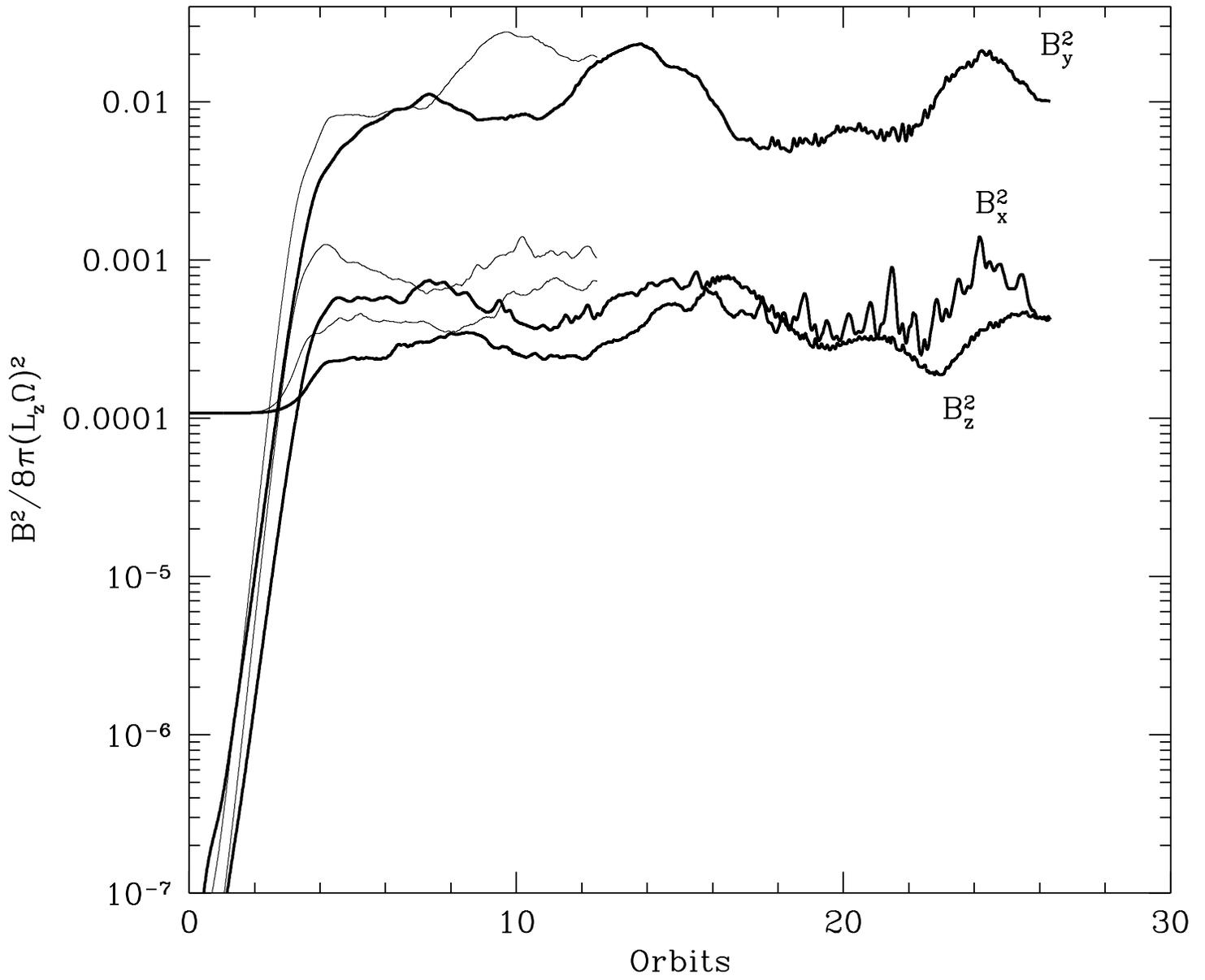
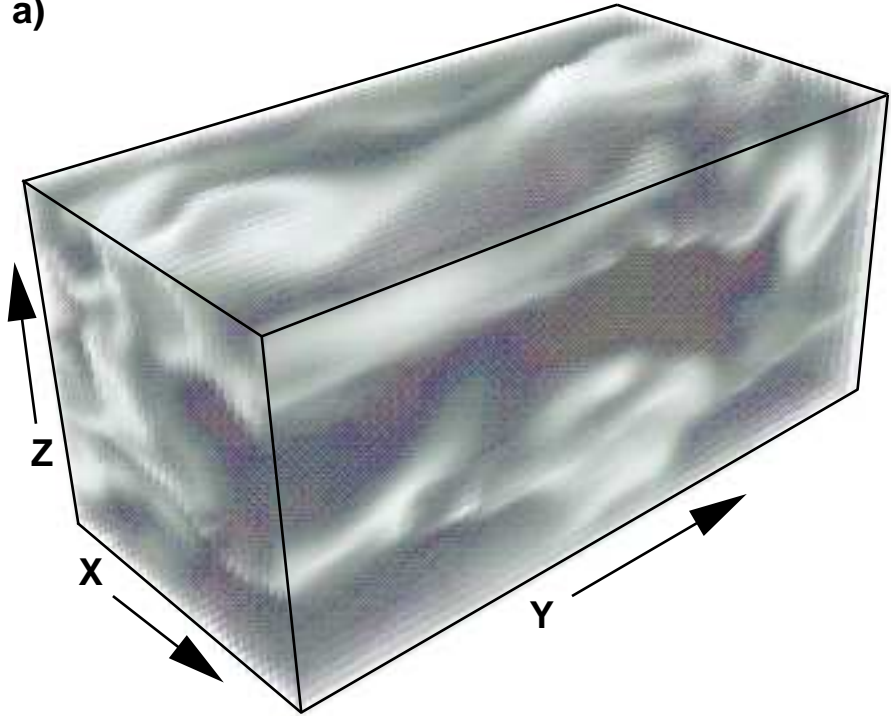
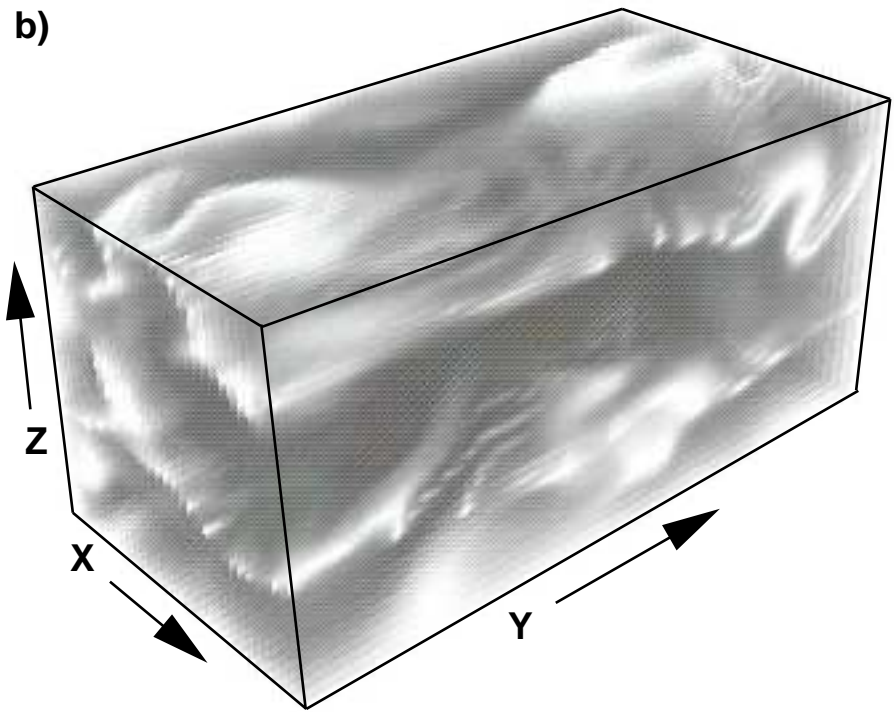
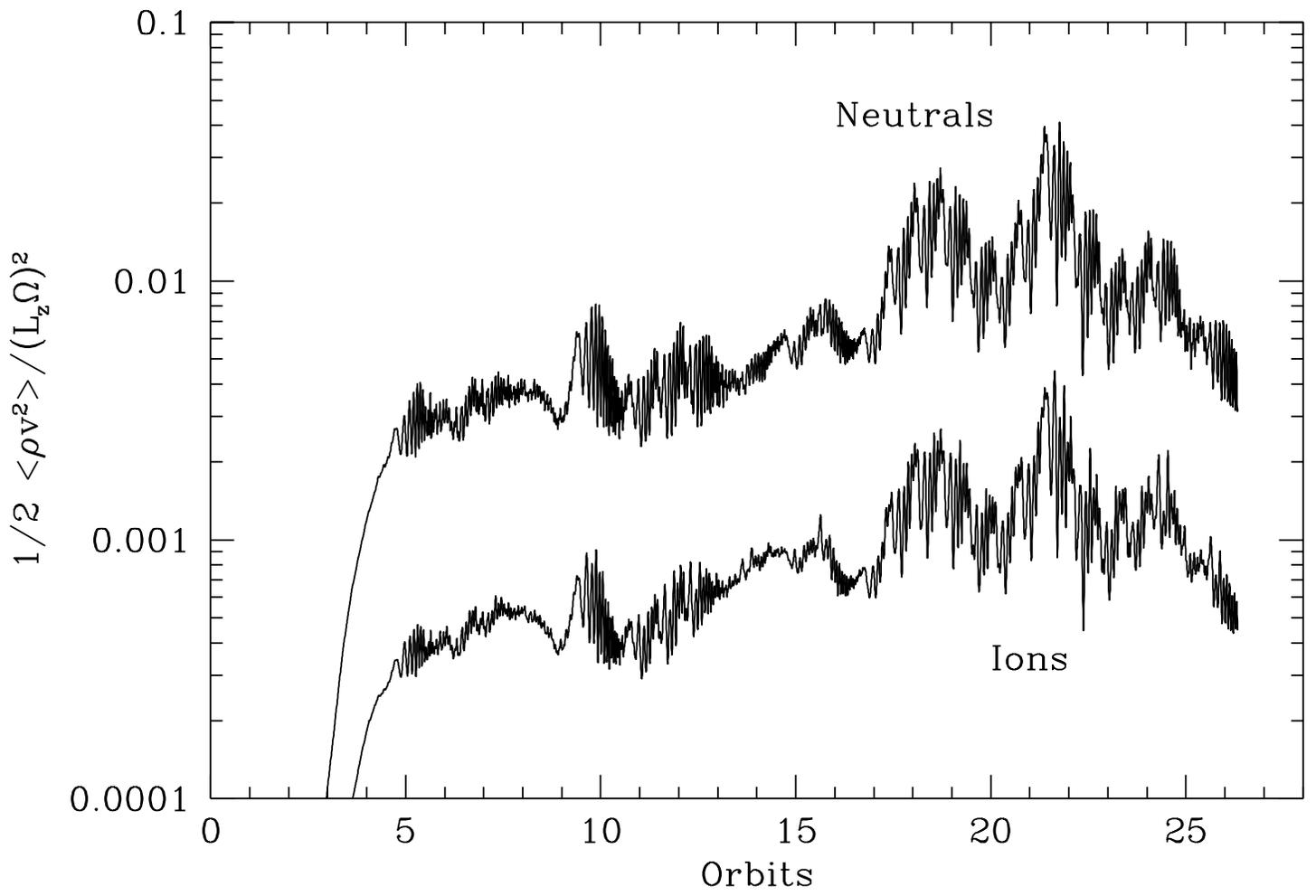
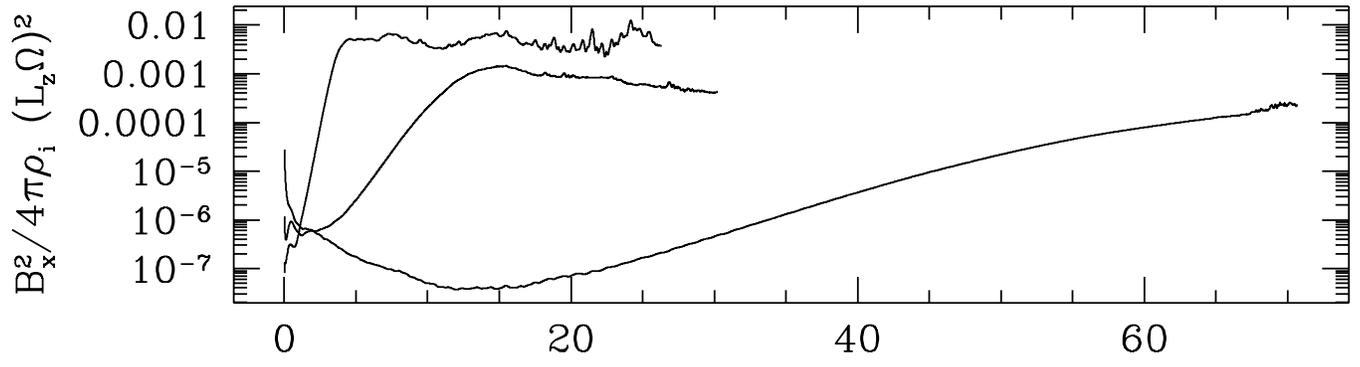
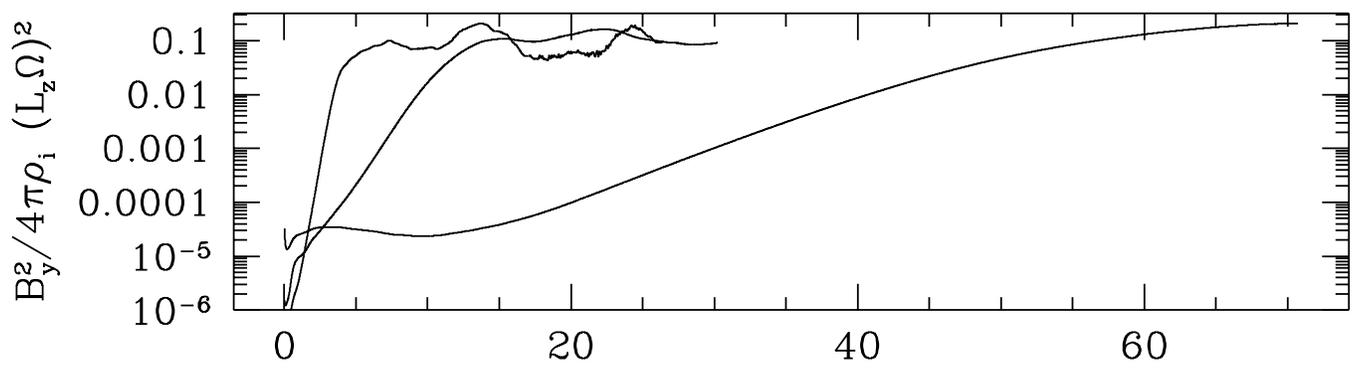
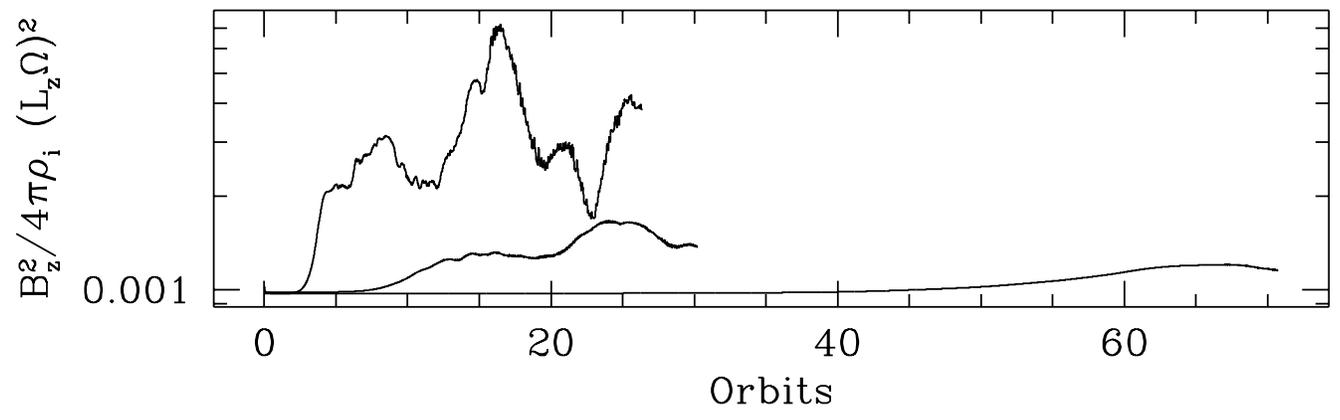
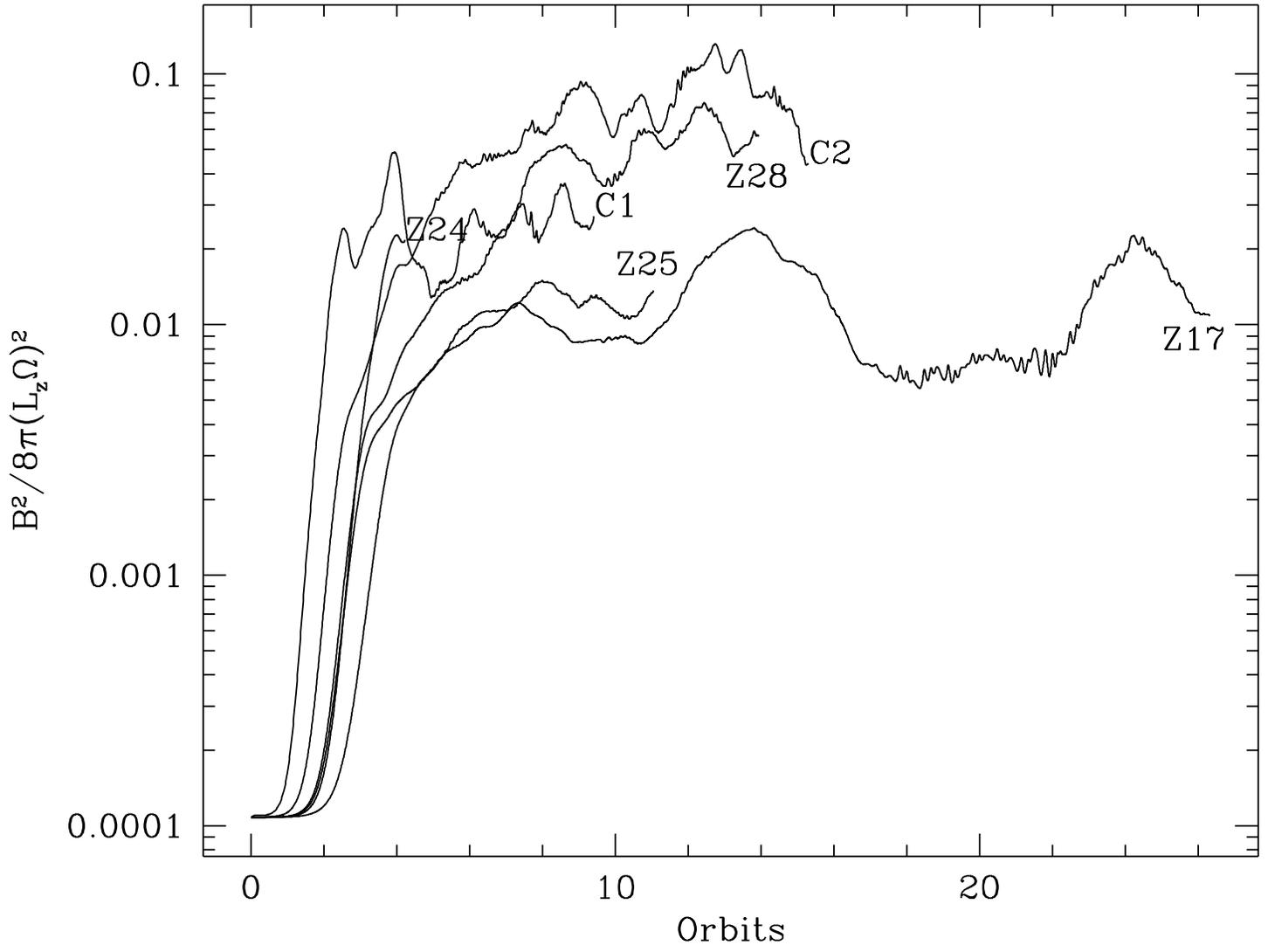
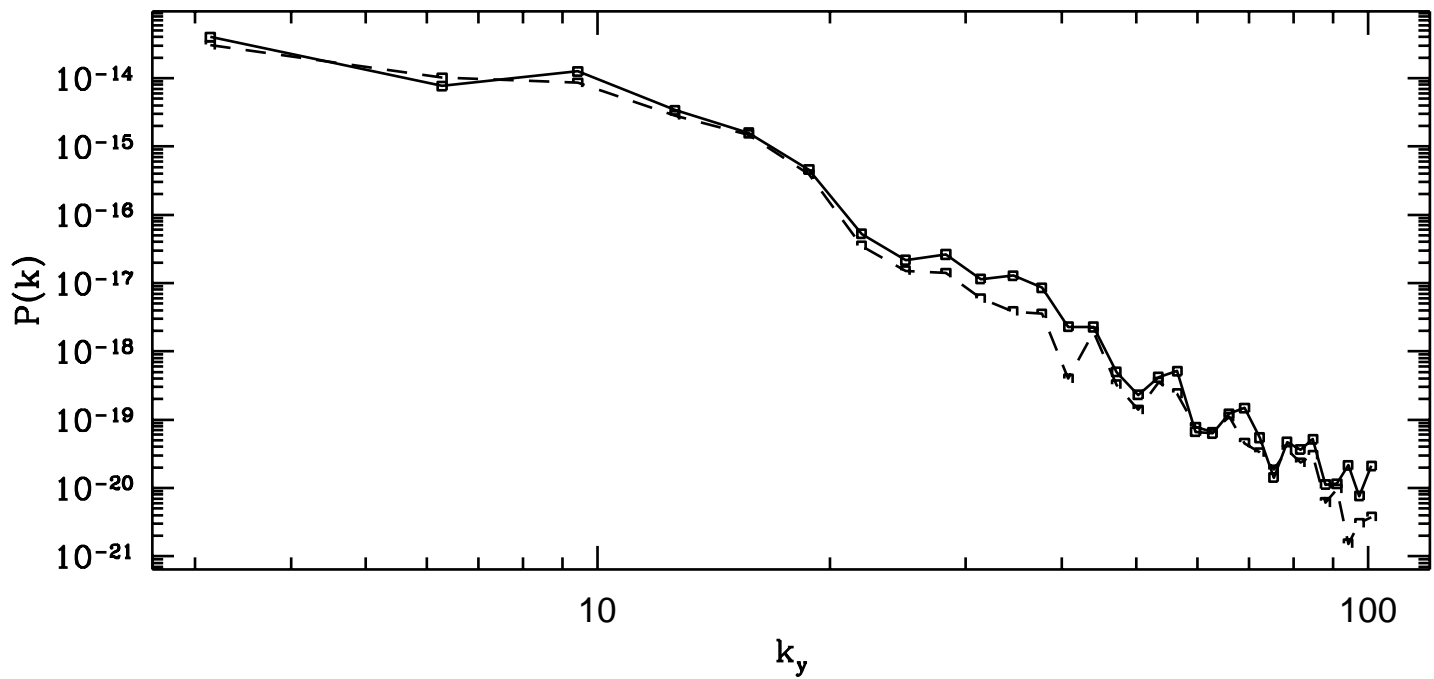
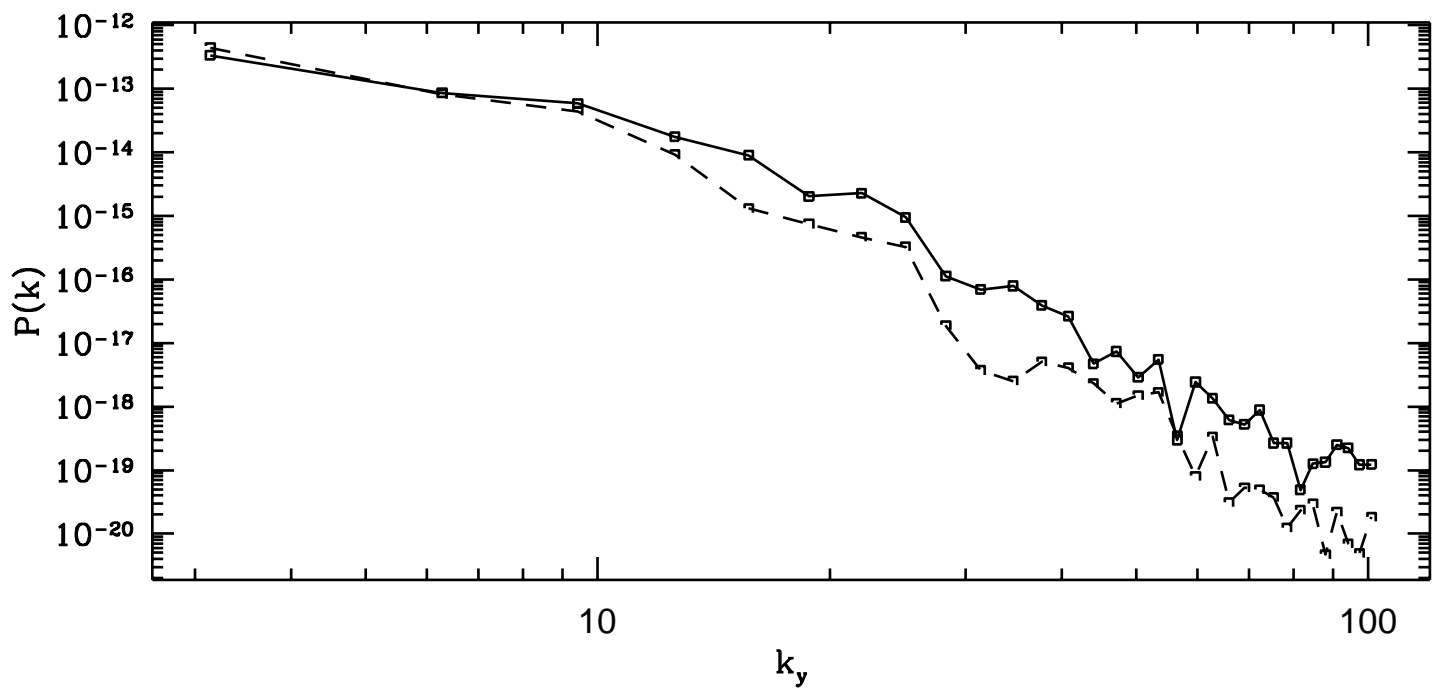
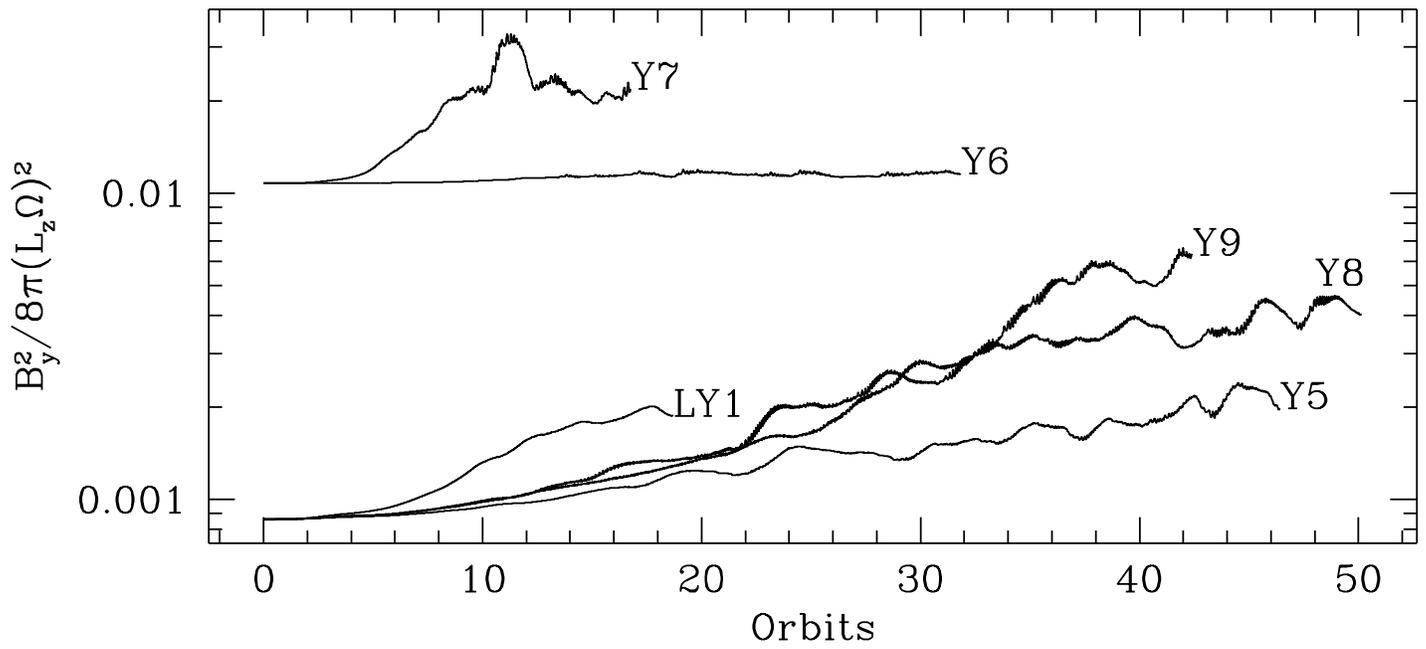
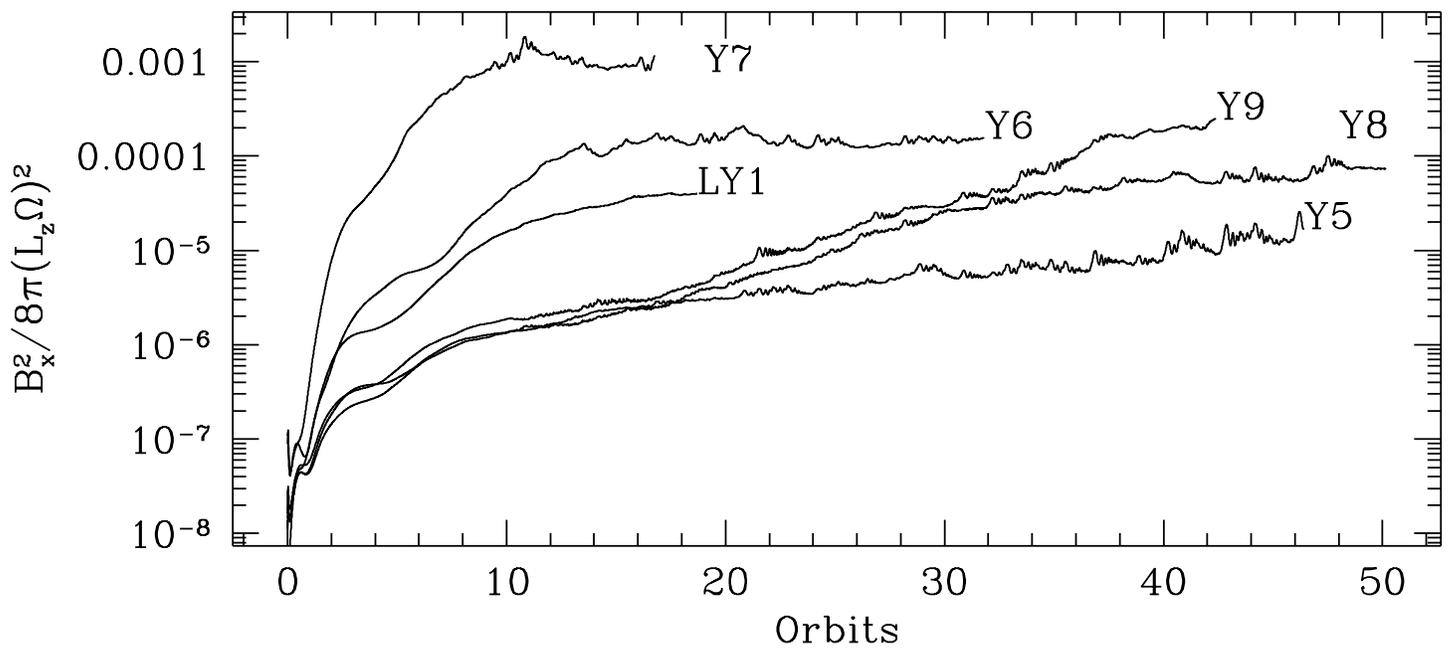
\end{document}